\documentclass[twocolumn]{autart}    

\newtheorem{definition}{Definition}
\newtheorem{remark}{Remark}
\newtheorem{ass}{Assumption}

\usepackage{hyperref}
\usepackage{algpseudocode}
\usepackage{algorithm}
\usepackage{amsmath,amsfonts,amssymb,bbm}     
\usepackage{graphicx}
\usepackage{csquotes}
\usepackage{bm}
\usepackage{subcaption}
\usepackage{xcolor}
\usepackage{amssymb}
\usepackage{amsmath}
\usepackage{mathtools}
\usepackage[shortlabels]{enumitem}
\usepackage{soul}
\usepackage{steinmetz}
\def\A{{\mathbf A}}
\def\B{{\mathbf B}}
\def\C{{\mathbf C}}

\def\C{{\mathbf C}}

\def\e{{\mathbf e}}
\def\F{{\mathbf F}}
\def\G{{\mathbf G}}
\def\H{{\mathbf H}}
\def\I{{\mathbf I}}

\def\L{{\mathbf L}}
\def\M{{\mathbf M}}

\def\S{{\mathbf S}}

\def\U{{\mathbf U}}
\def\u{{\mathbf u}}
\def\v{{\mathbf v}}
\def\V{{\mathbf V}}

\def\x{{\mathbf x}}
\def\y{{\mathbf y}}

\def\zero{{\mathbf 0}}

\def\1{{\mathbb I}}

\def\mA{{\mathcal A}}
\def\mB{{\mathcal B}}
\def\mC{{\mathcal C}}
\def\mE{{\mathcal E}}
\def\mF{{\mathcal F}}
\def\mG{{\mathcal G}}

\def\mJ{{\mathcal J}}

\def\mL{{\mathcal L}}

\def\mM{{\mathcal M}}

\def\mS{{\mathcal S}}
\def\mT{{\mathcal T}}
\def\mV{{\mathcal V}}

\def\we{{\widetilde e}}

\def\wx{ {\widetilde x}}

\def\hS{ {\widehat \S}}
\def\hL{ {\widehat \L}}

\def\mbE{{\mathbb E}}

\def\mbN{{\mathbb N}}
\def\mbR{{\mathbb R}}
\def\mbS{{\mathbb S}}
\def\mbZ{{\mathbb Z}}

\newcommand{\expectation}[1]{\mbE\left\{#1\right\}}

\def\tS{\mathbf {\tilde{\S}}}
\def\tL{\mathbf {\tilde{\L}}}
\def\ty{\mathbf {\tilde{\y}}}

\def\<{{\langle}}
\def\>{{\rangle}}

\def\bwe{ \mathbf{ {\widetilde e}}}
\def\bwH{ \mathbf{ {\widetilde H}}}
\def\bwL{ \mathbf{ {\widetilde L}}}
\def\bwS{ \mathbf{ {\widetilde S}}}

\newcommand{\eps}{\varepsilon}

\begin{document}

\begin{frontmatter}

\title{ Topology Identification under Spatially Correlated Noise\thanksref{footnoteinfo}} 

\thanks[footnoteinfo]{This work is supported by NSF through the project 
titled "RAPID: COVID-19 Transmission Network Reconstruction from Time-Series Data" 
under Award Number 2030096.}

\author[First]{Mishfad Shaikh Veedu}, \author[First]{Murti V. Salapaka} 

\address[First]{Department of Electrical and Computer Engineering, University of Minnesota, MN 55455, USA (e-mail: \{veedu002, murtis\}@ umn.edu).}


\begin{abstract}                          
This article addresses the problem of reconstructing the topology of a network of agents interacting via linear dynamics, while being excited by exogenous stochastic sources that are possibly correlated across the agents, from time-series measurements alone. It is shown, under the assumption that the correlations are affine in nature, such network of nodal interactions is equivalent to a network with added agents. The added agents are represented by nodes that are latent, where no corresponding time-series measurements are available; however, here all the exogenous excitements are spatially (that is, across agents)  uncorrelated. Generalizing affine correlations, it is shown that, under polynomial correlations, the latent nodes in the expanded networks can be excited by clusters of noise sources, where the clusters are uncorrelated with each other. The clusters can be replaced with a single noise source if the latent nodes are allowed to have non-linear interactions. Finally, using the sparse plus low-rank matrix decomposition of the imaginary part of the inverse power spectral density matrix (IPSDM) of the time-series data, the topology of the network is reconstructed. Under non conservative assumptions, the correlation graph of the noise sources is retrieved.
\end{abstract}

\end{frontmatter}

\section{Introduction}
Networks and graphical models provide convenient tools for effective representations of complex high dimensional multi-agent systems. Such a representation is useful in applications including power grids \cite{patel2017distributed,wood2013power}, meteorology \cite{climate}, neuroscience \cite{bower2012book}, and finance \cite{carfi2008financial}. Knowledge of the network interaction structure (also known as network topology) helps in understanding, predicting, and in many applications, controlling/identifying the system behavior \cite{Dreef_VandenHof,Bazanella_2020,Salapaka_signal_selection,Ramaswamy,ShenglingShi_VandenHof,VANDENHOF20132994}. In applications such as power grid, finance, and meteorological system, it is either difficult or impossible to intervene and affect the system. Hence, inferring the network properties by passive means such as time-series measurements is of great interest in these applications. 
An example is unveiling the correlation structure between the stocks in the stock market from daily share prices \cite{carfi2008financial}, which is useful in predicting the market behavior.

Learning the conditional independence relation between variables from time-series measurements is an active research field among machine learning (ML), probabilistic graphical model (PGM), and statistics communities \cite{Elena,cpw,joshin_online,joshin_2,Quinn_DIG,Yanning_top_id}, where the system modules are considered as random variables. However, such studies fail to capture dynamic dependencies between the entities in a system, which are prevalent in most of the aforementioned applications. For dynamical systems, autoregressive (AR) models that are excited by exogenous Gaussian noise sources, which are independent across time and variables, are explored in \cite{Alpago_ContLett_2018,alpago2021scalable,ARMA_identification_of_GM,songsiri10}. Here, the graph topology captured the sparsity pattern of the regressor coefficient matrices, which characterized the conditional independence between the variables. It was shown that the sparsity pattern of the inverse power spectral density matrix (IPSDM), also known as concentration matrix, identifies the conditional independence relation between the variables (see \cite{ARMA_identification_of_GM,songsiri10}). As shown in \cite{materassi_tac12}, the conditional independence structure between the variables is equivalent to the moral-graph structure of the underlying directed graph. For multi-agent systems with linear dynamical interaction, excited by wide sense stationary (WSS) noise sources that are mutually uncorrelated across the agents, moral graph reconstruction using Wiener projection has gained popularity in the last decade \cite{INNOCENTI_mat_Wiener_automatica,Innocenti_mat_TAC_2010,materassi_tac12}. Here, the moral-graph of the underlying linear dynamic influence model (LDIM) is recovered from the magnitude response of Wiener coefficients, obtained from time-series measurements. For a wide range of applications, the spurious connections in the moral-graph can be identified--returning the true topology--by observing the phase response of the Wiener coefficients \cite{TALUKDAR_physics}. Furthermore, for systems with strictly causal dynamical dependencies, Granger causality based algorithms can unveil the exact cause-effect nature of the interactions; thus recovering the exact parent-child relationship of the underlying graph \cite{materassi_GC,Dimovska_mat_tac20,materassi_tac12}. 

In many networks, time-series at only a subset of nodes is available. The nodes where the time-series measurements are not available form the latent/confound/hidden nodes. Topology reconstruction is more challenging in the presence of latent nodes as additional spurious connections due to confounded effects are formed when applying the aforementioned techniques. AR model identification of the independent Gaussian time-series discussed above, in the presence of latent nodes, was studied in \cite{AR3,Ciccone_TAC_2019,Ciccone_TAC_2020,Falconi_CDC_2020,Falconi_2021,AR1,zorzi_AR_hidden,ZORZI_Chiuso}. Here, the primary goal is to eliminate the spurious connections due to latent nodes and retrieve the original conditional independence structure from the observed time-series. 
In applications such as power grid, there is a need to retrieve the complete topology of the network, including that of the latent nodes. Such problems are studied for bidirected tree \cite{Talukdar_ACC18}, poly-forest \cite{materassi_blindlearning}, and poly-tree \cite{Salapaka_polytrees,materassi_polytree} networks excited by WSS noise sources that are uncorrelated across the agents. Recently, an approach to reconstruct the complete topology of a general linear dynamical network with WSS noise was provided in \cite{doddi2019topology}. The work in \cite{doddi2019topology} can be considered as a generalization of AR model identification with latent nodes, in asymptotic time-series regime. However, one major caveat of the aforementioned literature on graphical models is that the results fail if the exogenous noise sources are spatially correlated, i.e. if the noise is correlated across the agents/variables. Prior works have studied the systems with spatially correlated noise sources \cite{Fonken_VandenHof,Salapaka_signal_selection,Ramaswamy}; however, these studies assume the knowledge of topology of the network. In a related work, \cite{Rajagopal_CDC_21_top_corr} studied topology estimation under spatially correlated noise sources and used this estimated topology in local module identification. 

This article studies the problem of topology identification of the LDIMs that allow spatially \emph{correlated} noise sources, similar to the problem in \cite{Rajagopal_CDC_21_top_corr}. However, this article provides an alternate treatment, where the noise correlations are transformed to latent nodes. This transformation enables one to gain additional insights and apply techniques from topology identification with latent nodes to solve the problem.


The first major result of this article is to provide a transformation that converts an LDIM without latent nodes, but excited by spatially correlated exogenous noise sources, to LDIMs with latent nodes. Here, the latent nodes are characterized by the maximal cliques in the correlation graph, the undirected graph that represents the spatial correlation structure. It is shown that, under affine correlation assumption, that is, when the correlated noise sources are related in affine way (Assumption \ref{ass:affine_corre}), there exist transformed LDIMs with latent nodes, where all the nodes are excited by spatially uncorrelated noise sources. A key feature of the transformation is that the correlations are completely captured using the latent nodes, while the original topology remains unaltered. The original moral-graph/topology is the same as the moral-graph/topology among the observed nodes in the altered graph. Thus, this transformed problem is shown to be equivalent to topology identification of networks with latent nodes. Consequently, any of the aforementioned techniques for the networks with latent node can be applied on the transformed LDIM to reconstruct the original moral graph/topology.


Next, relaxing the affine correlation assumption, polynomial correlation is considered (Assumption \ref{ass:poly_corre}); here, the focus is on noise sources with distributions that are symmetric around the mean. It is shown that, in this scenario, the transformed dynamical model can be excited by clusters of noise sources, where the clusters are uncorrelated. However, the noise sources inside the clusters can be correlated. Using the sparse$+$low-rank decomposition of IPSDM from~\cite{doddi2019topology}, the true topology of the network along with the correlation structure is reconstructed, from the IPSDM of the original LDIM, without any additional information, if the network satisfies a necessary and sufficient condition. Notice that, the results discussed here are applicable to the networks with static random variables and AR models also, when the exogenous noise sources are spatially correlated.


The article is organized as follows. Section \ref{sec:systemmodel} introduces the system model, including LDIM, and the essential definitions. Section \ref{sec:IPSD_top_reconstruction} discusses IPSDM based topology reconstruction. Section \ref{sec:G_ctoG_t-trans} describes the correlation graph to latent nodes transformation and some major results. Section \ref{sec:poly_correl} discusses transformation of LDIMs with polynomial correlation to LDIMs with latent nodes. Section \ref{sec:topid_MD} explains the sparse$+$low-rank decomposition technique and how the topology can be reconstructed without the knowledge of correlation graph. Simulation results are provided in Section \ref{sec:sim}. Finally, Section \ref{sec:conclusion} concludes the article.

\textit{Notations:} 
Bold capital letters, $\A$, denotes  matrices; $A_{ij}$ and $(\A)_{ij}$ represents $(i,j)^{\text{th}}$ entry of $\A$; Bold small letters, $\v$, denotes vectors; $v_i$ indicates $i^{\text{th}}$ entry of $\v$; $[n]:=\{1,\dots,n\}$;
The subscript $o$ denotes the observed nodes index set $[n]$ and the subscript $h$ indicates the latent nodes index set $\{n+1,\dots, n+L\}$,  For example, $\x_o:=\{x_1,\dots,x_n\}$ and $\x_h:=\{x_{n+1},\dots,x_{n+L}\}$; $\Phi_\x(z):=[(\Phi_\x(z))_{ij}],~z \in \mathbb C,~ |z|=1$ denote power spectral density matrix, where $(\Phi_\x)_{ij}$ is cross power spectral density between the time-series at nodes $i$ and $j$; $\mF$ denotes the set of real rational single input single output transfer functions that are analytic on unit circle; $\mA$ denotes the set of rationally related zero mean jointly wide sense stationary (JWSS) scalar stochastic process; ${\mathcal Z}(.)$ denotes bilateral $z$-transform; $\mbS^n$ denotes the space of all skew symmetric $n \times n$ matrices; $\mathbb N$ denotes the set of natural numbers, $\{0,1,2,\dots\}$ ; For a set $\mS$, $|\mS|$ denotes the cardinality of the set. $\|\A\|_*$ denotes the nuclear norm and $\|\A\|_1$ denote sum of  absolute values of entries of $\A$. $\phase{a}$ denotes the phase of the complex number $a$. 

\section{System model}
\label{sec:systemmodel}
Consider a network of $n$ interconnected nodes, where node $i$ is equipped with time-series measurements, $(\widetilde{x}_i(t))_{t\in \mathbb{Z}}$, $i\in[n]$. The network interaction is described by,
\begin{align}\label{eq:Corr_TF_model}
	\x(z) &= \H(z)\x(z)+\e(z),
	\end{align}
where $\x(z)=[x_1(z), x_2(z),\dots,x_n(z)]$, $\x_i(z) = \mathcal{Z}[{\wx_i}]$. $\e(z)=[e_1(z), e_2(z),\dots,e_n(z)]\in \mA^n$ are the exogenous noise sources with $\Phi_{e}(z)$ is non-singular, but possibly non-diagonal. $\H(z)$ is the weighted adjacency matrix with $ H_{ii}=0$, for all $i \in [n]$. An LDIM is defined as the pair $(\H,\e)$, whose output process is given by \eqref{eq:Corr_TF_model}. The LDIM is well-posed if every entry of $(\I_n-\H(z))^{-1}$ is analytic on the unit circle, $|z|=1$ and topologically detectable if $\Phi_\e$ is positive definite.

Every LDIM has two associated graphs, viz: $\textbf{1}$) a directed graph, the linear dynamic influence graph (LDIG), $\mG(\mV,\mE)$, where $\mV=[n]$ and $\mE:=\{(i,j):H_{ji}\neq 0\}$ and $\textbf{2}$) an undirected graph, the correlation graph, $\mG_c(\mV,\mE_c)$, where $\mE_c:=\{(i,j):(\Phi_{e})_{ij} \neq 0, \ i \neq j\}$. Notice that, in a directed graph, if $(i,j)\in \mathcal{E}$ then there is a directed arrow from $i$ to $j$ in the graphical representation of $\mathcal{E}$ (see Fig. \ref{fig:moti}(a) for example).  Concisely, ``LDIG $(\H,\e)$" is used to denote the LDIG $\mG(\mV,\mE)$ corresponding to the LDIM $(\H,\e)$. For a directed graph $\mG(\mV,\mE)$, the parent, the children, and the spouse sets of node $i$ are  $Pa(i):=\{j:(j,i) \in \mE\}$, $Ch(i):=\{j:(i,j) \in \mE\}$, and $Sp(i):=\{j:j \in Pa(Ch(i))\}$  respectively. The topology of a directed graph $\mG(\mV,\mE)$, denoted $\mT(\mV,\mE)$, is the undirected graph obtained by removing the directions from every edge $(i,j) \in \mE$. Moral-graph (also called kin-graph) of $\mG(\mV,\mE)$, denoted $kin(\mG)$ is an undirected graph, $kin(\mG):=\mG(\mV,\mE')$, where $\mE':=\{(i,j): i \neq j, \ i \in \left( Sp(j) \bigcup Pa(j) \bigcup Ch(j) \right) \}$. A clique is a sub-graph of a given undirected graph where every pair of nodes in the sub-graph are adjacent. A maximal clique is a clique that is not a subset of a larger clique. For a correlation graph, $\mG_c(\mV,\mE_c)$, $q$ denotes the number of maximal cliques with clique size $>1$. 



The problem we address is described as follows.
\begin{prob}
\label{prob1}
(P1)
Consider a well-posed and topologically detectable LDIM,
$(\H,\e)$, where $\Phi_\e$ is allowed to be non-diagonal and its associated graphs $\mG(\mV,\mE)$ and $\mG_c(\mV,\mE_c)$ are unknown. Given the power spectral density matrix $\Phi_{\x}$, where $\x$ is given by \eqref{eq:Corr_TF_model}, reconstruct the topology of $\mG$.
\end{prob}
\section{IPSDM based topology reconstruction}
\label{sec:IPSD_top_reconstruction}
In this section, the IPSDM based topology reconstruction is presented.
In \cite{materassi_tac12}, the authors showed that, for any LDIM characterized by $\eqref{eq:Corr_TF_model}$, the availability of IPSDM, which can be written as
\begin{align}
\Phi_{\x}^{-1}=(\I_n-\H)^*\Phi_{e}^{-1} (\I_n-\H),
\end{align}is sufficient for reconstructing the moral-graph of $(\H,\e)$. That is, if $(\Phi_{\x}^{-1})_{ij}\neq 0$, then $i$ and $j$ are kins. However, an important assumption for the result to hold true is that $\Phi_{e}^{-1}$ is diagonal. If $\Phi_{e}^{-1}$ is non-diagonal, then the result does not hold in general. For $i \neq j$,

{ \footnotesize
\begin{align*}
(\Phi_{\x}^{-1})_{ij}&= (\Phi_{e}^{-1})_{ij}-(\H^*\Phi_{e}^{-1})_{ij}-(\Phi_{e}^{-1} \H)_{ij}
+(\H^*\Phi_{e}^{-1} \H)_{ij}
\\&=(\Phi_{e}^{-1})_{ij}-\sum_{k=1}^n (\H^*)_{ik}(\Phi_{e}^{-1})_{kj}-\sum_{k=1}^n (\Phi_{e}^{-1})_{ik} \H_{kj}
\\& \hspace{1cm }+\sum_{k=1}^n\sum_{l=1}^n (\H^*)_{ik}(\Phi_{e}^{-1})_{kl}\H_{lj},
\end{align*}}and any of the four terms can cause $(\Phi_{\x}^{-1})_{ij}\neq 0$, depending on $\Phi_\e^{-1}$. Hence, this technique cannot be applied directly to solve Problem \ref{prob1}. For example, consider the network $\mG$  given in Fig. \ref{fig:moti}(a) with $(\Phi_\e^{-1})_{13} \neq 0$. Then, $(\Phi_{\x}^{-1})_{23}\neq 0$ since $ \H^*_{21}(\Phi_\e^{-1})_{13} \neq 0$, which implies that the estimated topology has $(2,3)$ present, while $(2,3) \notin kin(\mG)$.

In the next section, the correlation graph, $\mG_c(\mV,\mE_c)$, is scrutinized and properties of $\mG_c(\mV,\mE_c)$ are evaluated as the first step in unveiling the true topology when noise sources admit correlations.
\section{Spatial Correlation to Latent Node Transformation}
\label{sec:G_ctoG_t-trans}


In this section, a  transformation of the LDIM, $(\H,\e)$, to an LDIM with latent nodes by exploiting the structural properties of the noise correlation is obtained. The transformation converts an LDIM without latent nodes and driven by spatially \emph{correlated} exogenous noise sources to an LDIM with latent nodes that are excited by spatially \emph{uncorrelated} exogenous noise sources. Assuming perfect knowledge of the noise correlation structure, the latent nodes and their children in the transformed LDIM are characterized. It is shown that although the transformation is not unique, the topology of the transformation is unique under affine correlation (Assumption \ref{ass:affine_corre}).


For $\bwe=[\bwe_o; \bwe_h]$ and $\bwH:=$ {\tiny $\left[\begin{matrix} \H&\F \\ \zero&\zero \end{matrix}\right]$}, $\H\in \mF^{n\times n},\ \F\in \mF^{n\times L},~L\in\mbN$, $(\bwH,\bwe)$ (or with a slight abuse of notation, $([\H,\F],\bwe)$) denotes an LDIM with $L$ latent nodes, where $\F_{ik}$ denotes the directed edge weight from latent node $k$ to observed node $i$ and $\H_{ik}$ denotes the directed edge weight from observed node $k$ to observed node $i$. Notice that the latent nodes considered in this article are strict parents and they do not have incoming edges. 



\subsection{Relation between Spatial Correlation and Latent Nodes}
\label{subsec:spatial_corre_to_latent_node}
Here, it is demonstrated that the spatially correlated exogenous noise sources can be viewed as the children of a latent node that is a common parent of the correlated sources. The idea is explained in the motivating example below. Towards this, let us define the following notion of equivalent networks.

\begin{definition}
\label{def:equiv_LDIMs}
Let $(\H^{(1)},\e^{(1)})$ and $(\H^{(2)},\e^{(2)})$ be two LDIMs and let $\x^{(1)}=(\I_n-\H^{(1)})^{-1}\e^{(1)}$ and $\x^{(2)}=(\I_n-\H^{(2)})^{-1}\e^{(2)}$ respectively. Then, the LDIMs $(\H^{(1)},\e^{(1)})$ and $(\H^{(2)},\e^{(2)})$ are said to be equivalent, denoted $(\H^{(1)},\e^{(1)}) \equiv (\H^{(2)},\e^{(2)})$ if and only if $\Phi_{\x^{(1)}}=\Phi_{\x^{(2)}}$.
\end{definition}


\begin{figure}
\centering
\begin{subfigure}[t]{0.4\linewidth}
\includegraphics[trim=250 220 240 200,clip, width=\textwidth]{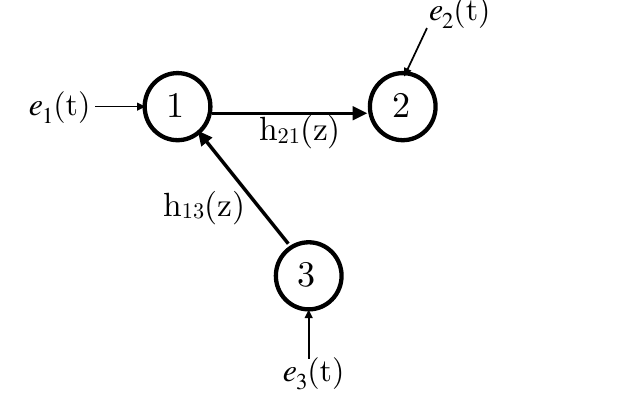}
\caption{Network with correlated noise sources, ($\Phi_\e$ non-diagonal).}
\label{fig:moti_corre}
\end{subfigure}
~
\begin{subfigure}[t]{0.4\linewidth}
\includegraphics[trim=250 170 240 200,clip, width=\textwidth]{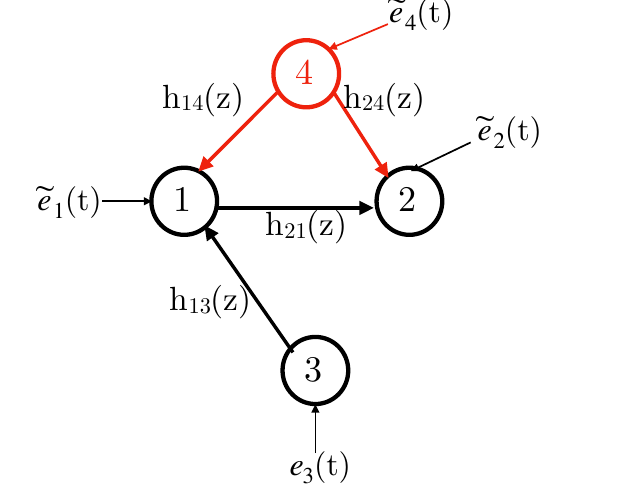}
\caption{Transformed model with latent node.} 
\label{fig:moti_transformed}
\end{subfigure}
\caption{Transformation of an LDIM with correlated noise sources to a network with uncorrelated noise sources in the presence of a latent node.} \label{fig:moti}
\end{figure}

Consider the network of three nodes in Fig. \ref{fig:moti}(a). Suppose the noise sources $e_1$ and $e_2$ are correlated, that is, $\Phi_{e_1 e_2} \neq 0$. Consider the network in Fig. \ref{fig:moti}(b), where $\widetilde{e}_1, \we_2, \we_4$, and $e_3$ are jointly uncorrelated and node $4$ is a latent node. Note that $e_3$ is same in both the LDIMs while $\we_1,\we_2$ are different from $e_1,e_2$. From the LDIM of Fig. \ref{fig:moti}(a), it follows that
\begin{align}
\label{eq:gen_graph_1}
    x_1&=e_1+h_{13}x_3 \ \& \ 
x_2=e_2+h_{21}x_1.
\end{align}

From the LDIM of Fig. \ref{fig:moti}(b), it follows that
\begin{align}
\begin{split}
x_1&=\we_1+h_{14}x_4 +h_{13}x_3,\\
\label{eq:gen_graph_2}    x_2&=\we_2+h_{24}x_4+h_{21}x_1,\ \& \
x_4=\we_4.
\end{split}
\end{align}
Comparing \eqref{eq:gen_graph_1} and \eqref{eq:gen_graph_2}, it is evident that if $\we_1, \ \we_2$, and $\we_4$ are such that $e_1=\we_1+h_{14}\we_4$ and $e_2=\we_2+h_{24}\we_4$, then the time-series obtained from both the LDIMs are identical. That is, the LDIMs are equivalent.

For a given LDIM, $(\H,\e)$, with correlated exogenous noise sources, one can define a space of equivalent LDIMs with uncorrelated noise sources that provide the same time-series. The following definition captures this space of ``LDIMs with latent nodes and uncorrelated noise sources" that are equivalent to the original LDIM with correlated noise sources. 


\begin{ass}
\label{ass:affine_corre}
In LDIM $(\H,\e)$, defined by \eqref{eq:Corr_TF_model}, the exogenous noise processes $e_i,e_j \in \mA$ are correlated only via affine interactions, i.e., $\Phi_{e_ie_j} \neq 0$  if and only if there exists an affine transform, $f(x)=a+bx$, $a\in \mA,~b \in \mF,~b\neq 0$ such that either $e_i=f(e_j)$ or $e_j=f(e_i)$.
\end{ass}

\begin{definition}
\label{def:mL}
For any LDIM, $(\H,\e)$,  with $\Phi_\e$ non-diagonal, and satisfying Assumption \ref{ass:affine_corre},
{\footnotesize\begin{align}
\begin{split}
\label{eq:def_mL}
\mL(\H,\e)&:=\{(\bwH,\bwe): \e=\bwe_o+\F\bwe_h,\ \bwH= {\tiny \left[\begin{matrix} \H&\F \\ \zero&\zero \end{matrix}\right]},\ \bwe=[\bwe_o,\bwe_h]^T,\\ 
&\hspace{-10pt}\ \bwe\in \mA^{n+L},\ \F\in \mF^{n\times L}, \ L \in \mbN, \ \Phi_{\bwe}\text{ diagonal}\},
\end{split}
\end{align}}
is the space of all $\mL$-transformations for a given $(\H,\e)$. 
\end{definition}
\begin{remark}
In \eqref{def:mL}, the number of latent nodes, $L$, is not fixed a priori.
\end{remark}
Based on the definition of $\mL(\H,\e)$ in \eqref{eq:def_mL}, LDIM for $([\H,\F],\bwe)$ can be rewritten as
\begin{align}
\label{eq:modified_TF_model_sysid}
	\x(z) &= \H(z)\x(z)+\G(z)\bwe(z),
\end{align}
where $\bwe(z)=[\we_1(z),\dots,\we_n(z),\dots,\we_{n+L}(z)]$ are mutually uncorrelated and $\G=\begin{bmatrix}\I_n,   \F\end{bmatrix}$. 

\begin{figure}
\centering
\begin{subfigure}[t]{0.2\textwidth}
\includegraphics[trim=230 250 340 220,clip, width=\textwidth]{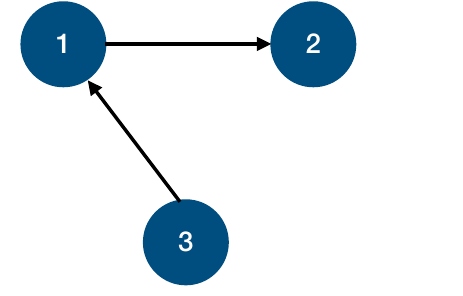}
\caption{Original LDIG, $\mG(\mV,\mE)$.}
\label{fig:L_trans_gen}
\end{subfigure}
~
\begin{subfigure}[t]{0.2\textwidth}
\includegraphics[trim=230 250 340 220,clip, width=\textwidth]{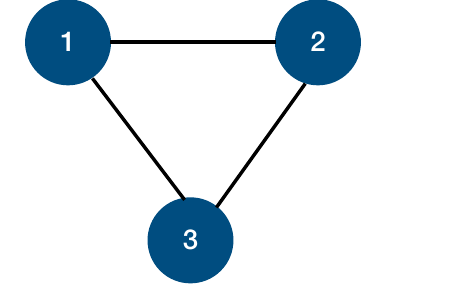}
\caption{Correlation graph, $\mG_c(\mV,\mE_c)$, of $\mG(\mV,\mE)$.} 
\label{fig:L_trans_corre}
\end{subfigure}

\begin{subfigure}[t]{0.2\textwidth}
\includegraphics[trim=30 258 530 120,clip, width=\textwidth]{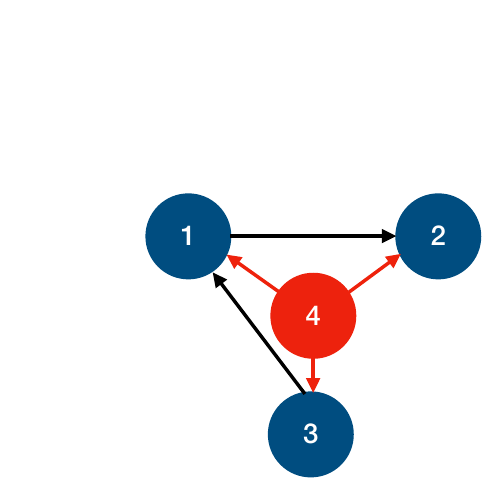}
\caption{Example transformed graph in $\mL(\H,\e)$ with one latent node.} 
\label{fig:L_trans_one_HN}
\end{subfigure}
\begin{subfigure}[t]{0.2\textwidth}
\includegraphics[trim=500 258 30 120,clip, width=\textwidth]{mot_transformed_graphs.pdf}
\caption{Example transformed graph in $\mL(\H,\e)$ with three latent nodes.}
\label{fig:L_trans_two_HN}
\end{subfigure}

\caption{Demonstration of $\mL(\H,\e)$; noise sources are not shown. The exogenous noises are uncorrelated in Fig. \ref{fig:L_trans_demo}(c) and \ref{fig:L_trans_demo}(d).} \label{fig:L_trans_demo}
\end{figure}

An LDIM, $(\bwH,\bwe) \in \mL(\H,\e)$, is called a \emph{transformed} LDIM of $(\H,\e)$ in this article. Intuitively, $\mL$-transformation completely assigns the spatial correlation component present in the original noise source, $\e$, to latent nodes, without altering the LDIG, $\mG(\mV,\mE)$, similar to the discussion on the aforementioned motivating example. For the LDIM in Fig. \ref{fig:L_trans_demo}(a) with correlation graph in Fig. \ref{fig:L_trans_demo}(b), Fig.  \ref{fig:L_trans_demo}(c), and Fig. \ref{fig:L_trans_demo}(d) show some examples of the LDIGs that belong to $\mL(\H,\e)$. Thus, $\mL$-transformation returns a larger network with latent nodes. Notice that the latent nodes present in an LDIG of $(\bwH,\bwe) \in \mL(\H,\e)$, are \emph{strict parents}, whose interaction with the true nodes are characterized by $\F$. 

The following theorem formalizes the relation between the correlation graph and the latent nodes in the transformed higher dimensional LDIMs. In particular, it shows that two noise sources $e_i$ and $e_j$ are correlated if and only if there is a latent node, which is a common parent of both the nodes $i$ and $j$ in the transformed LDIM.

\begin{lem}
\label{lem:clique_G_c-vs-latent_node}
	Let $(\H,\e)$ be an LDIM defined by \eqref{eq:Corr_TF_model} that satisfies Assumption \ref{ass:affine_corre}, and let $\mG_c(\mV,\mE_c)$ be the correlation graph of $\{e_i\}_{i=1}^n$. 
	Then, for every distinct $i,j \in [n]$, $(i,j)\in \mE_c$ if and only if every LDIG $([\H,\F],\bwe) \in  \mL(\H,\e)$ contains a latent node $h$ such that $h \in Pa(i) \cap Pa(j)$.
\end{lem}

\textbf{Proof:} Refer Appendix \ref{app:clique_G_c-vs-latent_node} \hfill \hfill \qed

Thus, $e_i$ and $e_j$ in the original LDIM are correlated if and only if for every $\mL$-transformed LDIM there exists a $k$ such that $F_{ik}\neq0$ and $F_{jk}\neq0$.

The following lemma shows that the number of latent nodes, $L$, present in any $([\H,\F],\bwe) \in \mL(\H,\e)$ is at least $q$, the number of maximal cliques with clique size greater than one in $\mG_c$.
\begin{lem}
\label{lem:Q_leq_L}
Let $(\H,\e)$ be an LDIM and let $\mG_c(\mV,\mE_c)$ be the correlation graph of the exogenous noise sources, $\e$. Then, the number, $L$, of latent nodes present in any  $([\H,\F],\bwe) \in \mL(\H,\e)$ satisfies $L \geq q$, where $q$ is the number of maximal cliques with clique size $>1$ in $\mG_c$.
\end{lem}
\textbf{Proof:} Refer Appendix \ref{app:lem_Q_leq_L}.\hfill \hfill \qed
\begin{remark}
Let $\mG_c(\mV_c,\mE_c)$ be the correlation graph of the exogenous noise sources. Then, for any given maximal clique $\mG^\ell(\mV^\ell,\mE^\ell)\subseteq\mG_c(\mV_c,\mE_c)$ and for any $k_\ell \geq 1$, there exists a transformed LDIG $([\H,\F],\bwe) \in \mL(\H,\e)$ with $k_\ell$ number of latent nodes such that the set $\mV^\ell$ is equal to the children of the latent nodes. For $|\mV^\ell|=3$, Fig. \ref{fig:L_trans_demo} shows examples with $k_l=1,3$ (see proof of Lemma \ref{lem:Q_leq_L} for details). 
\end{remark}

As shown in Lemma \ref{lem:Q_leq_L} and Fig. \ref{fig:L_trans_demo}, the LDIMs in the characterizing space $\mL(\H,\e)$ of the equivalent LDIMs can have an arbitrary number of latent nodes, which leads to multiple transformed representations with varying number of latent nodes. The following definition restricts the number of latent nodes present in the equivalent transformed LDIGs and provides a minimal transformed representation--representation with minimal number of nodes--of the true LDIG. 

\begin{definition}
Define $\mL_q(\H,\e)$ to be the set of all LDIMs in $\mL(\H,\e)$ with the number of latent nodes equal to the number of maximal cliques with clique size $>1$ present in $\mG_c$.
\end{definition}



The following lemma shows that there exist a unique latent node in every transformed LDIG $([\H,\F],\bwe)\in \mL_q(\H,\e)$, corresponding to each maximal clique.

\begin{thm}
\label{thm:L=Q}
Let $(\H,\e)$ be an LDIM that satisfies Assumption \ref{ass:affine_corre} and let $\mG_c(\mV,\mE_c)$ be the correlation graph of the exogenous noise sources,$e$. Suppose $\mG_c$ has $q$ maximal cliques $C_1,\dots,C_q$. Consider any LDIG $([\H,\F],\bwe) \in \mL_q(\H,\e)$ with $q$ latent nodes, $h_1,\dots,h_q$. Then, for a maximal clique $C_i$ in $\mG_c$, there exists a unique latent node $h\in\{h_1,\dots,h_q\} $ such that $Ch(h)=C_i$.
\end{thm}
\textbf{Proof:} Refer Appendix \ref{app:lem_L=Q}.\hfill \hfill \qed

\begin{remark}
    Notice that the existence of the unique latent node is true only for the space $\mL_q(\H,\e)$. Such a unique node might not exist for the transformed representations in $\mL(\H,\e)$. In other words, Theorem 4 identifies the minimal set of latent nodes required to explain the data.
\end{remark}
\subsection{Uniqueness of the topology}
Here, the LDIMs in $\mL_q(\H,\e)$ from the previous section is studied more carefully. In topology reconstruction, only the support structure of the transfer functions matter. The following proposition shows that the support structure is unique.

\begin{prop}
\label{prop:topology_uniqueness}
Let $(\H,\e)$ be an LDIM that satisfies Assumption \ref{ass:affine_corre}. The topology of every LDIG $([\H,\F],\bwe) \in  \mL_q(\H,\e)$ is the same.

\end{prop}

\textbf{Proof:} Refer Appendix \ref{app:proof_prop_top_uniqueness}.\hfill \hfill \qed

Based on the above results, the following transformation of Problem (P1) is formulated.

\begin{prob} (P2)
Consider an LDIM $(\H,\e)$ defined by \eqref{eq:Corr_TF_model} that satisfies Assumption \ref{ass:affine_corre}, and $\Phi_\e$ allowed to be non-diagonal. Let $([\H,\F],\bwe) \in \mL_q(\H,\e)$ be an LDIM with LDIG given by $\mG_t(\mV_t,\mE_t)$. Suppose that the time-series data among the observed nodes of $([\H,\F],\bwe)$ is given. Then, reconstruct the topology among the observed nodes of $\mG_t$.
\end{prob}

\begin{remark}
The time series data among the observed nodes of the transformed LDIM is the same as the time series obtained from the original LDIM.
\end{remark}
\begin{remark}
\label{remark:P1=P2}
The problems (P1) and (P2) are equivalent. That is, the topology reconstructed in both the problems are the same, because topology among the observed nodes of any element of $\mL(\H,\e)$ is same as the topology of $(\H,\e)$ (see Definition \ref{def:mL} and proof of Proposition \ref{prop:topology_uniqueness}).
\end{remark}
\medskip

Therefore, instead of reconstructing the topology of $(\H,\e)$ with $\Phi_e$ non-diagonal, it is sufficient to reconstruct the topology among the observed nodes for one of the LDIMs $ ([\H,\F],\bwe)\in \mL_q(\H,\e)$, which has $\Phi_{\bwe}$ diagonal.

\section{Polynomial Correlation}
\label{sec:poly_correl}
The results in the previous section assumed that the correlations between the exogenous noise sources are affine in nature. In this section, a generalization of the affine correlation is addressed. It is shown that the noise sources that are correlated via non-affine, but a polynomial, interaction can be characterized using latent nodes with non-linear interaction dynamics. By lifting the processes to a higher dimension, the non-linearity is converted to linear interactions.

The following definitions are useful for the presentation in this section.

\begin{definition} \cite{IVA}
Let $x=(x_1,\dots,x_m)$. A monomial in $x_1,\dots,x_m$ is a product of the form  $x^\alpha:=x_1^{\alpha_1}x_2^{\alpha_2} \dots x_m^{\alpha_m}$, where $\alpha \in \mathbb N^m$. A polynomial $P$ in $x$ with coefficient in a field (or a commutative ring), $\mathbb F$, is a finite linear combination of monomials, i.e.,
\[P(x):=\sum_{\alpha} a_\alpha x^\alpha,~ a_\alpha \in \mathbb F.  \]

\begin{enumerate}[(i)]
    \item The degree of a monomial $x^\alpha$ is $|\alpha|:=\sum_{i=1}^n \alpha_i$.
    \item The total degree of $P\neq 0$ is the maximum of $|\alpha|$ such that $a_\alpha \neq 0$.
\end{enumerate}

\end{definition}

\begin{definition}
For any $\v \in \mA^m,$ define the list of monomials of total degree at most $p$, $$\mM(\v,p):=[f_0(\v),f_1(\v), \dots, f_p(\v)]^T,$$ where $f_k(\v)$ lists all the $k$-degree monomials, $f_k(\v):=\{\v_1^{\alpha_1} \v_2^{\alpha_2}\dots \v_m^{\alpha_m}\mid \sum_{i=1}^m\alpha_i=k\}$. For example, when $m=2$, $f_0(\v)=1$, $f_1(\v)=\{\v_1,\v_2\}$, and $f_2(\v)=\{\v_1^2,\v_1\v_2,\v_2^2\}$. In general, the total number of monomials having degree $k$ is given by $\binom{m+k-1}{k}$, where $\binom{n}{k}=\frac{n!}{k!(n-k)!}$. The total number of monomials up to total degree $p$ is $M=${\tiny$\displaystyle \sum_{k=0}^p\binom{m+k-1}{k}$}. For $m=2$ and $p=3$, $\mM(\v,3)=[1,\v_1,\v_2,\v_1^2,\v_1\v_2,\v_2^2,\v_1^3,\v_1^2\v_2,\v_1\v_2^2,\v_2^3]^T$.
\end{definition}

\subsection{Characterization of Polynomial Correlations}
\label{subsec:char_poly_correl}

Suppose $e_1=\sum_{|\alpha|\leq M}a_{\alpha,1} \v^\alpha$ and $e_2=\sum_{|\alpha|\leq M}a_{\alpha,2} \v^\alpha$, where $a_{\alpha,1},a_{\alpha,2} \in \mF$ and $\v \in \mA^m$, $m,M\in \mathbb N$, with $\Phi_\v$ diagonal. Let $\y=\mM(\v,p)$ be the vector obtained by concatenating $\v^\alpha$, $|\alpha|\leq p$. Letting $e_1=\A_1\y$ and $e_2=\A_2\y$, we have $\Phi_{e_1e_2}=\A_1\Phi_\y \A_2^*$,  where $\A_i$ is the vector obtained by concatenating $a_{\alpha,i} $ .

Notice that $\y$ is lifting of $\v$ into a higher dimensional space of  polynomials. In the following, a discussion on the structure of $\Phi_\y$ is provided, with the help of examples.


\subsection{Example: Lifting of a zero mean Gaussian Process}
\label{subsec:IID_Gauss}
To illustrate lifting of noise processes to higher dimension, consider independent and identically distributed (IID) Gaussian process (GP). It is shown that, under lifting, the power spectral density matrix (PSDM) is block diagonal. 

Consider an IID, GP $\{\v(k), k\in \mbZ \mid \v(k)\sim N(0,\sigma^2\I_m)\}$. Then, $\expectation{\v_i\v_j}=\expectation{\v_i} \expectation{\v_j}=0$. It can be shown that \cite{papoulis} \begin{equation}
\label{eq:moments_Gaussian}
    \expectation{\v_i^p}=\left\{\begin{array}{cc}
    0 &  \text{ if $p$ is odd,}\\
    \sigma^p(p-1)!! & \text{ if $p$ is even,}
\end{array}\right.
\end{equation}
where $p!!:=(p-1)(p-3)\dots 3.1$ denotes the double factorial. Consider $m=2$ and let $\y=[y_1,\dots,y_{10}]:=\mM(\v,3)$ $=[1,~\v_1,~\v_2,~\v_1^2,~\v_1\v_2,~\v_2^2,~\v_1^3,~\v_1^2\v_2,~\v_1\v_2^2,~\v_2^3]^T$. That is, $\y$ lists all the monomials of $\v_1,\v_2$ with degree $\leq 3$. 


Notice that $\expectation{\v_1^i\v_2^j}=\expectation{\v_1^i} \expectation{\v_2^j}\neq 0$ if and only if both $i$ and $j$ are even. Then, $\expectation{y_2y_5}=\expectation{\v_1^2\v_2}=0$. Straight forward computation shows that $\expectation{y_2y_k}\neq 0$ only for $k=2,7,9$. Similarly, $\expectation{y_7y_k} \neq 0$ if and only if $k=7,9,2$, and $\expectation{y_9y_k} \neq 0$ if and only if $k=7,9,2$. Notice that $k=2,7,9$ corresponds to the terms with odd power on $x_1$ and even power on $x_2$.
Repeating the same for every $\expectation{y_iy_k}$, $1\leq i,k \leq 10$, one can show that, after appropriate rearrangement of rows and columns, the covariance matrix and the PSDM of $\y$ forms a block diagonal matrix, given by \eqref{eq:psd_blockdiagonal_ytilde}. Here, $\ty=[y_1,y_4,y_6,y_2,y_7,y_9, y_3,y_8,y_{10},y_5]$.
{\tiny
\begin{align}
\label{eq:psd_blockdiagonal_ytilde}
\Phi_{\ty}= \left[\begin{matrix}
\Phi_{11} & \Phi_{14} & \Phi_{16} & 0 & 0 & 0& 0& 0& 0&0\\
\Phi_{41} & \Phi_{44} & \Phi_{46} & 0 & 0 & 0& 0& 0& 0&0\\
\Phi_{61} & \Phi_{46} & \Phi_{66} & 0 & 0 & 0& 0& 0& 0&0\\
 0 & 0 & 0& \Phi_{22} & \Phi_{27} & \Phi_{29} & 0& 0& 0&0\\
 0 & 0 & 0& \Phi_{72} & \Phi_{77} & \Phi_{79} & 0& 0& 0&0\\
 0 & 0 & 0&\Phi_{92} & \Phi_{97} & \Phi_{99} & 0& 0& 0&0\\
0 & 0 & 0& 0& 0& 0 & \Phi_{55}& 0& 0&0\\
0 & 0 & 0& 0& 0& 0 & 0& \Phi_{33}& \Phi_{38}&\Phi_{310}\\
0 & 0 & 0& 0& 0& 0 & 0& \Phi_{83}& \Phi_{88}&\Phi_{810}\\
0 & 0 & 0& 0& 0& 0 & 0& \Phi_{103}& \Phi_{108}&\Phi_{1010}\\
\end{matrix}\right]
\end{align}}

It is worth mentioning that since the Gaussian process is zero mean IID, the auto-correlation function $R_{\v_i\v_j}(k)=R_{\v_i\v_j}(0)\delta(k)=\expectation{\v_i(0)\v_j(0)}\delta(k)$, where $\delta$ is Kronecker delta and thus $\Phi_{\ty}(z)=R_{\ty}(0)$, $\forall |z|=1$ is white (same for all the frequencies). 


\begin{figure}
\centering
\begin{subfigure}[t]{.4\linewidth}
\includegraphics[trim=30 350 590 130,clip, width=\textwidth]{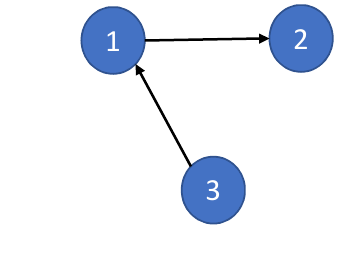}
\caption{Original LDIG, $\mG(\mV,\mE)$}
\label{fig:poly_moti_LDIG}
\end{subfigure}
\begin{subfigure}[t]{.4\linewidth}
\includegraphics[trim=30 135 590 330,clip, width=\textwidth]{polynomial_correlation_lift_example.pdf}
\caption{A correlation graph of $\mG(\mV,\mE)$}
\label{fig:poly_moti_G_c}
\end{subfigure}

\begin{subfigure}[t]{.8\linewidth}
\includegraphics[trim=230 150 160 130,clip, width=\textwidth]{polynomial_correlation_lift_example.pdf}
\caption{An example transformed LDIG of sub figures above with $m=2$, $p=3$, and $n=3$ with $2^m=4$ clusters. Here, $e_1=\we_1+F_{12}y_2+F_{17}y_7+F_{19}y_9$, $e_2=\we_2+F_{22}y_2+F_{27}y_7+F_{29}y_9+F_{21}y_1+F_{24}y_4+F_{26}y_6$, and $e_3=\we_3+F_{31}y_1+F_{34}y_4+F_{36}y_6$.}
\label{fig:poly_moti_corre}
\end{subfigure}
~
\begin{subfigure}[t]{.8\linewidth}
\includegraphics[trim=250 170 140 170,clip, width=\textwidth]{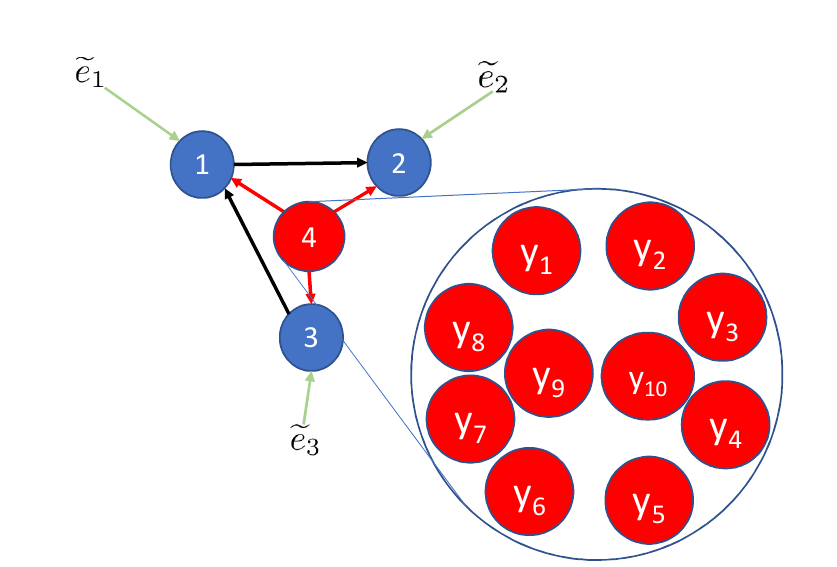}
\caption{Transformed graph of the above LDIG when all the three nodes are correlated. The latent node $4$ will capture the interactions due to the nodes $y_1,\dots,y_{10}$.} 
\label{fig:poly_moti_transformed}
\end{subfigure}
\caption{Transformation of an LDIM with correlated noise sources to a network with uncorrelated noise sources in the presence of a latent node.} \label{fig:poly_moti}
\end{figure}

The following proposition shows this for general $m,p$.
\begin{prop}
\label{prop:Gaussian_block_diagonal}
Consider the Gaussian IID process $\{\v(k), k\in \mbZ \mid \v(k)\sim N(0,\sigma^2\I_m)\}$. Let $\y=\mM(\v,p)$. Then, there exists a $\ty$, a permutation of $\y$, such that $\Phi_\ty$ is block diagonal with $2^m$ non-zero blocks.
\end{prop}
\textbf{Proof: } Refer Appendix \ref{app:prop_Gaussian_block_diagonal}.\hfill \hfill \qed





Based on this example, one can extend the result to symmetric zero mean WSS processes. Symmetric distributions are the distributions with probability density functions symmetric around mean, for example, Gaussian distribution.

\begin{definition}
\label{def:symmetric_distri}
A probability distribution is said to be symmetric around mean if and only if it's density function, $f$, satisfies $f(\mu-x)=f(\mu+x)$ for every $x\in \mbR$, where $\mu$ is the mean of the distribution.
\end{definition}

\begin{prop}
\label{prop:WSS_blockdiag}
Consider a zero mean WSS process, $ \v \in  \mA^m$ with symmetric distribution and $\Phi_\v$ diagonal. Let $\y=\mM(\v,p)$. Then, there exists a $\ty$, a permutation of $\y$ such that $\Phi_\ty$ is block diagonal with $2^m$ non-zero blocks.
\end{prop}
\textbf{Proof:} Proof is similar to the IID GP case. For symmetric distributions, odd moments are zero, since odd moments are the integral of odd functions. \hfill \hfill \qed

As shown in the proof of Proposition \ref{prop:Gaussian_block_diagonal}, the monomial nodes corresponding to a given block diagonal (i.e., the same odd-even pattern) can be grouped into a cluster. Fig. \ref{fig:poly_moti}(c) shows such a clustering with $m=2$ and $p=3$. The red nodes are the lifted processes in the higher dimension (the monomial nodes $y_1,\dots,y_{10}$). The nodes inside a given cluster (shown by blue ellipse) are correlated with each other and the nodes from different clusters are not correlated with each other; this is a result of $\Phi_{\ty}$ being block-diagonal. Here, $e_1$ and $e_2$ are correlated via the cluster of $y_2,~y_7,$ and $y_9$, whereas $e_2$ and $e_3$ are correlated via the cluster of $y_1,~y_4,$ and $y_6$. Notice the following important distinction from the LDIMs in Fig. \ref{fig:moti}. Here, it is sufficient for the observed nodes $1,2$ to be connected to one of the latent nodes in a given cluster. It is not necessary for the nodes to have a common parent like the case in Fig. \ref{fig:moti}. The common parent property from Section \ref{sec:G_ctoG_t-trans} is replaced with a common ancestral cluster here.

\subsection{Transformation of Polynomial Correlation to Latent Nodes}

Based on the aforementioned discussion, relaxation of Assumption \ref{ass:affine_corre} and extension of the LDIM transformation results from Section \ref{sec:G_ctoG_t-trans} is presented here. It is shown that in order to relax Assumption \ref{ass:affine_corre}, non-linear interactions are required between the latent nodes and the observed nodes. The following assumption is a generalization of Assumption \ref{ass:affine_corre}.

\begin{ass}
\label{ass:poly_corre}
In LDIM $(\H,\e)$, defined by \eqref{eq:Corr_TF_model}, the noise processes $e_i,e_j \in \mA$ are correlated if and only if there exist polynomials $P_1,P_2$ and $\bwe=[\v^T,\we_1,\we_2]^T$, $\v \in \mA^{m}$, $e_1,e_2 \in \mA$, with $\Phi_{\bwe}$ diagonal, $m\in \mbN$, such that $e_1=\we_1+P_1(\v)$ and  $e_2=\we_2+P_2(\v)$, where $P_i(\v)=\sum_{|\alpha|\leq M}a_{\alpha,i} \v^\alpha$, $a_{\alpha,i} \in \mF$, for some $M\in \mbN$.
\end{ass}
\begin{remark}
To be precise, the extension of Assumption \ref{ass:affine_corre} is given by $y=P(x)$, where $x \in \mA$ and $P(x)=\sum_{\alpha}a_{\alpha} x^\alpha$, $a_{\alpha} \in \mF$ is a polynomial of degree less than or equal to $M$.
Any $x\in \mA$ can be written as $x=\we_x+bv$, where $\we_x,v \in \mA$, $b\in \mF$, and $\we_x$ uncorrelated with $v$. Then, $y=P(\we_x+bv)$, which is a special case of Assumption \ref{ass:poly_corre}.
\end{remark}


\begin{definition}
\label{def:mL_poly}
For any LDIM, $(\H,\e)$,  with $\Phi_\e$ non-diagonal, and satisfying Assumption \ref{ass:poly_corre}, and $,~p>1$
{\footnotesize
\begin{align}
\nonumber
\label{eq:def_mL_p}
\mL^{(p)}(\H,\e)&:=\{(\bwH,\y):\e=\bwe_o+\F \y,~\bwH={\tiny \left[\begin{matrix} \H&\F \\ \zero&\zero \end{matrix}\right]},~\F \in \mF^{n\times M},\ \\ 
\nonumber
&\hspace{-0.9cm}\ \y=\mM(\v,p), ~\bwe_o\in \mA^{n},~ \v\in \mA^{m},~m \in \mbN,~  \bwe=[\bwe_o,\v]^T,\\ 
&\hspace{0cm} M= \sum_{k=0}^p{\tiny\binom{m+k-1}{k}},~ \Phi_{\bwe}\text{ diagonal}\}
\end{align}}
\normalfont is the space of all $\mL_p$-transformations for a given $(\H,\e)$. The matrix $\F$ is obtained by concatenating $\F_\alpha\in \mF^{n\times 1}$. This is done by first listing all the $\F_\alpha$ corresponding to ``degree one" monomials in lexicographic order, then ``degree $2$" in lexicographic order, etc. That is,{\footnotesize$\F:=[\F_{b_1},\dots,\F_{b_L},\F_{b_{11}},\F_{b_{12}},\dots, \F_{b_{1L}},\F_{b_{21}},\F_{b_{22}},\dots],$} where $\F_{b_{i_1\dots i_k}}$ denotes the column vector of $\F$ corresponding to $\alpha=b_{i_1}+\dots+b_{i_k}$, $b_i:$ canonical basis. 
For $m=2,~p=3$ from Section \ref{subsec:IID_Gauss}, $\F=[\F_{b_1},\F_{b_2},\F_{b_{11}},\F_{b_{12}},\F_{b_{22}},\F_{b_{111}},\F_{b_{112}},\F_{b_{122}},\F_{b_{222}}]$. Here, $\F_{b_{2}}=\F_{(0,1)}, ~\F_{b_{12}}=\F_{(1,1)},~\F_{b_{222}}=\F_{(0,3)}$.

\end{definition}
With this new ``polynomial lifting" definition, one can transform a given LDIM with correlated noise sources to a transformed LDIM with latent nodes. As shown in the following lemma, transformation of correlations to uncorrelated latent nodes from Section \ref{sec:G_ctoG_t-trans} is replaced with uncorrelated latent clusters here. For any cluster $c$, $Ch(c):=\{Ch(i): i\in c\}$, that is, $Ch(c)$ denotes the union of the children of the nodes present in cluster $c$.

\begin{lem}
\label{lem:poly_clique_G_c-vs-latent_node}
	Let $(\H,\e)$ be an LDIM that satisfies Assumption \ref{ass:poly_corre}, and let $\mG_c(\mV,\mE_c)$ be the correlation graph of the exogenous noise sources, $\{e_i\}_{i=1}^n$. 

Then, for every distinct $i,j \in [n]$, $(i,j)\in \mE_c$ if and only if for every LDIG $(\bwH,\y) \in  \mL^{(p)}(\H,\e)$, there exists a cluster $c$ such that $i,j\in Ch(c)$.
\end{lem}

\textbf{Proof:} Refer Appendix \ref{app:poly_clique_G_c-vs-latent_node}.\hfill \hfill \qed

The following theorem shows that a subgraph, $\mG^\ell(\mV^\ell,\mE^\ell)$, of the correlation graph, $\mG_c(\mV,\mE_c)$, forms a maximal clique in $\mG_c$ if and only if for any transformed LDIG in $\mL^{(p)}(\H,\e)$, the set of nodes in $\mV^l$ is equal to the set of the children of some latent cluster.

\begin{thm}
\label{thm:prop_poly_clique-vs_lat-nodes}
Let $(\H,\e)$ be an LDIM defined by \eqref{eq:Corr_TF_model} which satisfies Assumption \ref{ass:poly_corre},and let $\mG_c(\mV,\mE_c)$ be the correlation graph of the exogenous noise sources, $\e$. Suppose that $\mG^\ell(\mV^\ell,\mE^\ell)\subseteq \mG_c(\mV,\mE_c)$ is a maximal clique with $|\mV^\ell|>1$. Then, for any LDIM $(\bwH,\y) \in  \mL^{(p)}(\H,\e)$, there exist latent clusters $C_1^\ell,\dots,C^\ell_{k_\ell}$ in the LDIG of $(\bwH,\y)$ such that
\begin{align}
\label{eq:V^l_condition_poly}
\mV^\ell&=\bigcup_{i=1}^{k_\ell}Ch(C^\ell_i) \text{ and }
\mE^\ell= \bigcup_{i=1}^{k_\ell}\mE_{\ell,i},
\end{align} 
where  $\mE_{\ell,i}:=\{(k,j):k,j \in Ch(c_i^\ell)\}$. In particular, for any latent cluster $C$ in the LDIG of $(\bwH,\y)$, $Ch(C) $ forms a clique in $\mG_c$.
\end{thm}
\textbf{Proof:} See Appendix \ref{app:thm_poly_clique-vs_lat-nodes}.\hfill \hfill \qed

Consider the LDIG shown in Fig. \ref{fig:poly_moti}(a) with correlation graph given by Fig. \ref{fig:poly_moti}(b). As shown in Fig. \ref{fig:poly_moti}(d), if all the three nodes are correlated, one can transform this LDIM to an LDIM with latent node $4$. However, here node $4$ should capture the interactions from the latent nodes $y_1-y_{10}$. Therefore, during reconstruction, one should accommodate non-linear interaction between the latent node and the observed nodes.

In the next section, we describe a way to perform the reconstruction, when $\mG_c$ is unavailable.

\section{Moral-Graph Reconstruction by Matrix Decomposition}

\label{sec:topid_MD}
In the previous sections, frameworks that convert a network with spatially correlated noise to a network with latent nodes were studied. In this section, a technique is provided to reconstruct the topology of the transformed LDIMs using the sparse plus low-rank matrix decomposition of the IPSDM obtained from the observed time-series. Note that this result does not require any extra information other than the IPSDM obtained from the true LDIM.

\subsection{Topology Reconstruction under Affine Correlation}
Here, reconstruction under affine correlation is discussed. Recall from Definition \ref{def:mL} that $\e=\bwe_o+\F\bwe_h$. Then, PSDM of $\e$, $\Phi_{e}$, can be written as \cite{materassi_tac12}:
$\Phi_{e}=\Phi_{\bwe_o}+\Phi_{\bwe_o\bwe_h}\F^*+\F\Phi_{\bwe_h\bwe_o}+\F \Phi_{\bwe_h} \F^*$
$=\Phi_{\bwe_o}+\F \Phi_{\bwe_h} \F^*$,
where the second equality follows because $\bwe_h$ and $\bwe_o$ are uncorrelated with mean zero. 
Then, IPSDM of the observed nodes in $([\H,\F],\bwe) \in \mL(\H,\e)$,
{\begin{align}
\nonumber
\Phi_{o}^{-1}&=(\I_n-\H)^*(\Phi_{\we_o}+\F \Phi_{\we_h} \F^*)^{-1} (\I_n-\H)\\
\label{eq:correlation_to_A-B}
&\stackrel{(a)}{=}\A-\B,
\end{align}}
where $\A=(\I_n-\H)^*\Phi_{\we_o}^{-1}(\I_n-\H)$ and  $\B=(\I_n-\H)^*\Phi_{\we_o}^{-1}\F(\Phi_{\we_h}^{-1}+ \F^*\Phi_{\we_o}^{-1}\F)^{-1}\F^*\Phi_{\we_o}^{-1}(\I_n-\H)
$. Equality (a) follows from the matrix inversion lemma \cite{matrix_analysis}. \eqref{eq:correlation_to_A-B} can then be rewritten as:	{ \begin{align}	\label{eq:S_L_decomp}
	\Phi_{o}^{-1}=&\S+\L,\text{ where }\\
	\label{eq:S_def}
	\S&=(\I_n-\H^{*})\Phi_{\we_o}^{-1}(\I_n-\H),\\
	\label{eq:L_def}
	\L &=-\Psi^{*}\Lambda^{-1}\Psi,
\end{align}} $\Psi=\F^{*}\Phi_{\we_o}^{-1}(\I_n-\H)$, and 	$\Lambda=\F^{*}\Phi_{\we_o}^{-1}\F+\Phi_{\we_h}^{-1}$, which is similar to the model in \cite{doddi2019topology}. If the moral-graph of the original LDIG is sparse and $M\ll n$, then $\S$ is sparse and $\L$ is low-rank. It was shown in \cite{materassi_tac12} that the support of $\S$ retrieves the moral graph of $\mG$. Furthermore, as shown in \cite{doddi2019topology}, under some assumptions that is applicable to a large class of problems, $(i,j)$-th entry of $\S$ is strictly real if and only if the edge $i-j$ is a strict spouse edge. Thus, it can be shown that, in a large class of problems, support of $\Im\{\S\}$  retrieves the exact topology of $\mG$. Following the approach in \cite{doddi2019topology}, we reconstruct the network topology from the sparse+low-rank decomposition of $\Im\{\Phi_o^{-1}(z)\}$, which is a skew-symmetric matrix.
For completeness, the essential theories and a relevant algorithm from \cite{doddi2019topology} is provided below in Section \ref{subsec:sufficient_condition}. The idea here is to decompose a given skew-symmetric matrix, i.e. $\Im\{\Phi_o^{-1}(z)\}$, into the sparse and low-rank components ($\Im\{\S(z)\}$ and $\Im\{\L(z)\}$ respectively), and then to reconstruct the moral graph/topology from $\Im\{\S(z)\}$, for some $|z|=1$.


The next subsection shows how the sparse low-rank decomposition is applicable in polynomial correlation setting also.

\subsection{Topology reconstruction under polynomial correlation}
Under polynomial correlation, recall from Definition \ref{def:mL_poly} that $\e(z)= \bwe_o(z)+ \F \ty$. Let $\x_h=\bwe_h =\ty$ and let $\x=[\x_o^T, \x_h^T]^T$. Then, one can write,

{\footnotesize\begin{align}
\label{eq:latent_model}
    \left[\begin{matrix}
    \x_o(z)\\ \x_h(z)
    \end{matrix}\right]=\begin{bmatrix}\H(z)&\F(z)\\ \mathbf{0}_{M \times n}&\mathbf{0}_{M \times M} \end{bmatrix} \begin{bmatrix}\mathbf{x}_o(z)\\\mathbf{x}_h(z)\end{bmatrix}+\begin{bmatrix}\mathbf{\we_o}(z)\\\mathbf{\we_h}(z)\end{bmatrix},
\end{align}}
where, $\F_{ij}$ captures the influence of $\ty_j$ on $x_i$ and $\Phi_{\ty}$ is block diagonal. 

 The topology of the sub-graph restricted to the observed nodes in the above LDIM and the true topology of the network are equivalent. Moreover, similar to \eqref{eq:correlation_to_A-B} (see \cite{doddi2019topology} for details), one can  obtain \eqref{eq:S_L_decomp} to \eqref{eq:L_def} exactly.

Here $\S(z),\L(z)\in \mathbb{C}^{n \times n}$, $\Psi \in \mathbb{C}^{M\times n}$, and $\Lambda \in \mathbb{C}^{M\times M}$. If $M\ll n$, then $\L$ is low-rank. Let $\mJ:=\{j \in [M]\mid \exists i\in [n]\text{ with } \F_{ij}\neq 0 \}$ be the set of monomials that has non zero contribution to $e_i,~i\in [n]$ and let $ L=|\mJ|$. Under this scenario, $\L$ is low-rank if $L\ll n$. Section \ref{subse:simu_poly} demonstrates an example on application of the sparse plus low-rank decomposition to reconstruct the topology under polynomial correlation.

\subsection{Low-rank$+$Sparse Matrix Decomposition}
\label{subsec:lowrank-sparse_decomp} Here, the following problem is considered: given a matrix $\C$ that is known to be sum of a sparse skew-symmetric matrix $\S$ and a low-rank skew-symmetric matrix $\L$, retrieve the sparse and low-rank components. The following optimization program modified from \cite{CSPW_siam_2011} is used to obtain the sparse low-rank decomposition, where $0\leq t\leq 1$ is a pre-selected penalty factor \cite{doddi2019topology}.
{\footnotesize\begin{align}
\label{eq:convex_opti_t}
\begin{split}
(\widehat{\S}_t,\widehat{\L}_t)=& \arg \min_{\hS,\hL} t \|\hS\|_1 + (1-t)\|\hL\|_{*}\\
& \text{ subject to } \hS+\hL=\C,\\
& \hspace{1.55cm} \hS^T=-\hS, \ \hL^T=-\hL,
\end{split}
\end{align}}
where $\C=\Im\{\Phi_{o}^{-1}(z)\}$, for some $z\in \mathbb C,\ |z|=1$. 

In the next subsection, a sufficient condition and an algorithm for the exact recovery of the sparse and the low-rank components from $\C$ using \eqref{eq:convex_opti_t} are provided.
\subsection{Sufficient Condition for Sparse Low-rank Matrix Decomposition}
\label{subsec:sufficient_condition}
In this subsection, a sufficient condition (proved in \cite{CSPW_siam_2011,doddi2019topology}) is provided to uniquely decompose a matrix as a sum of the sparse skew-symmetric and the low-rank skew-symmetric components. Furthermore, an algorithm is provided that utilizes this sufficient condition to retrieve the sparse and low-rank components.

The following definitions are used in the subsequent results.

{\small \begin{align}
\nonumber deg_{max}(\M)&:=\max\left(\max_{1\leq i\leq n} \left( \sum_{j=1}^n \mathbbm{1}_{\{\M_{ij}\neq 0\}}\right)\right.,\\
& \hspace{2cm}\max_{1\leq j\leq n} \left.\left( \sum_{i=1}^n \mathbbm{1}_{\{\M_{ij}\neq 0\}}\right)\right),\\
inc(\M)&:=\max_k \|\U \U^T e_k\|_2,
\end{align}}where $\U\Sigma\V^T$ is the compact singular value decomposition of $\M$ and $\|\cdot\|_2$ is the Euclidean norm of a vector.

The following is a sufficient condition that guarantees the unique decomposition of $\C$ (see \cite{CSPW_siam_2011,doddi2019topology} for details).
\begin{lem}
\label{lem:suff_cond}
Suppose that we are given a matrix $\C$, which is the sum of a sparse matrix $\tS \in \mbS^n$ and a low-rank matrix $\tL \in \mbS^n$. If $(\tS,\tL)$ satisfies
\begin{align}
\label{eq:transverse_intersection}
deg_{max}(\tS) inc(\tL)<\frac{1}{12},
\end{align}
then there exists a penalty factor $t \in [0,1]$ such that \eqref{eq:convex_opti_t} returns $(\widehat{\S}_t,\widehat{\L}_t)=(\tS,\tL)$.
\end{lem}

\begin{remark}
The results in \cite{CSPW_siam_2011,doddi2019topology} are proved for the optimization problem with the objective function $\gamma \|\hS\|_1 +\|\hL\|_{*}$, where $\gamma \geq 0$. However, the results hold for $\eqref{eq:convex_opti_t}$ as well since the problems are equivalent (via the map $t=\gamma/(1+\gamma)$)
\end{remark}

The sufficient condition \eqref{eq:transverse_intersection} roughly translates to $\tS$ being sparse and the number of maximal cliques, $M$, being small, with clique sizes not too small (see \cite{doddi2019topology} for more details).


The following metrics are used to measure the accuracy of the estimates $(\hS_t,\hL_t)$ in the optimization \eqref{eq:convex_opti_t}.
{\footnotesize
\begin{align}
\nonumber
tol_{t}&:=\frac{\|\widehat{\S}_{t}-\tS\|_F}{\|\tS\|_F}+\frac{\|\widehat{\L}_{t}-\tL\|_F}{\|\tL\|_F},\\
 \label{eq:diff_t}
{diff}_t&:=(\| \widehat{\S}_{t-\epsilon}-\widehat{\S}_{t} \|_F)+(\| \
\widehat{\L}_{t-\epsilon}-\widehat{\L}_{t} \|_F),
\end{align}}where $\|.\|_F$ denotes the Frobenius norm and $\epsilon>0$ is a sufficiently small fixed constant. Note that $tol_{t}$ requires the knowledge of the true matrices $\bwS$ and $\bwL$, whereas ${diff}_t$ does not.

The following Lemma is proved in \cite{doddi2019topology} and is applied in Algorithm \ref{alg:matrix_decomposition} later to retrieve the sparse and low-rank components. 
\begin{lem}
	\label{lem:diff_t}
	Suppose we are given a matrix $\C$, which is obtained by summing $\tS$ and $\tL$, where $\tS$ is a sparse skew-symmetric matrix and $\tL$ is a low-rank skew-symmetric matrix. If $\tS$ and $\tL$ satisfies $\deg_{\max}(\tS) inc(\tL)<1/12$, then there exist at least three regions where $diff_t=0$. In particular, there exists an interval $[t_1,t_2]\subset[0,1]$ with $0<t_1<t_2<1$ such that the solution of \eqref{eq:convex_opti_t} is $(\hS_t,\hL_t)=(\tS,\tL)$ for any $t \in [t_1,t_2]$.
\end{lem}

Following the procedure in \cite{doddi2019topology}, moral-graph/topology reconstruction from the imaginary part of the IPSDM is considered here.
The following are some assumptions from \cite{doddi2019topology}, assumed for the exact recovery of the topology. The details can be found in \cite{doddi2019topology}, and are skipped here. In the absence of Assumption 4, the reconstruction algorithm will detect some false positive edges, but none of the true edges are missed.
\begin{ass}
\label{ass:H_real_const}
For any $i\in [n]$, $k\in [n]$ $(k'\in [L])$, if $\H_{ik}(z)\neq0,$ $(F_{ik'}\neq 0)$ then $\Im\{\H_{ik}(z)\}\neq 0$ $(\Im\{\F_{ik'}(z)\}\neq 0)$, for all $z\in \mathbb C, ~|z|=1$.
\end{ass}
\begin{ass}
\label{ass:talukdar_genera}
For the LDIM in \eqref{eq:Corr_TF_model}, and $i,k,l \in \mV$, if $\H_{ki}(z) \neq 0$ and $\H_{kl}(z) \neq 0$, then $\phase{\H_{ki}(z)}=\phase{ \H_{kl}(z)}$.  \end{ass}
The following Lemma from \cite{doddi2019topology} reconstructs the exact topology of the LDIM from $\S$ in \eqref{eq:S_def}.
\begin{lem}
\label{lem:S_top_obs_rec}
Consider a well-posed and topologically detectable LDM $(\H,\e)$ described by \eqref{eq:Corr_TF_model} with the associated graph $\mG(\mV,\mE)$, and satisfying Assumption \ref{ass:talukdar_genera}. Let $\S(z)$ be given by \eqref{eq:S_def} and let $\widehat{\mE}_o:=\{(i,j):\Im\{\S_{ij}(z)\}\neq 0,~i<j\}$ and $\overline{\mE}_o:=\{(i,j) \mid (i,j)\in \mE \text{ or } (j,i) \in \mE,~ i<j\}$, for some $z\in \mathbb{C}$, $|z|=1$. Then, $\widehat{\mE}_o\subseteq \overline{\mE}_o$. Additionally, if the LDM satisfies Assumption \ref{ass:H_real_const}, then $\hat{\mathcal{E}}_o=\overline{\mathcal{E}}_o$ almost always. 
\end{lem}

The following is the main result of this section, which follows from Lemma \ref{lem:suff_cond} and Lemma \ref{lem:S_top_obs_rec}.

\begin{thm}
\label{thm:s+L_top}
Let $(\H,\e)$ be an LDIM with $\Phi_e$ non-diagonal that satisfy Assumption \ref{ass:H_real_const}-Assumption \ref{ass:talukdar_genera}. Suppose that there exists an $(\H, \F,\bwe) \in \mL_q(\H,\e)$ such that for $\S$ and $\L$ in \eqref{eq:S_L_decomp},  $\deg_{\max}(\Im\{\S\}) inc(\Im\{\L\})<1/12$. Then, the true topology of $(\H,\e)$ can be reconstructed by solving the optimization problem $\eqref{eq:convex_opti_t}$ with $\C=\Im\{\Phi^{-1}_o(z)\}$ for some $z=e^{i\omega},~\omega \in (0,2\pi]$. In particular, $\mT(\mV,\mE)=supp(\hS_t)$, for appropriately selected $t$.
\end{thm}
\textbf{Proof:}
Refer Appendix \ref{app:proof_thm_S+L_top}.\hfill \hfill \qed

Additionally, applying the algorithms in \cite{doddi2019topology}, the correlation graph also can be reconstructed, if the LDIM satisfies the following assumption, as shown in simulation results.
\begin{ass}
\label{ass:latent5hop}
For every distinct latent nodes $k_h,k_h'$ in the transformed dynamic graph, the distance between $k_h$ and $k_h'$ is at least four hops.
\end{ass}


\begin{algorithm}
\caption{Matrix decomposition}
\textbf{Input:}$\Phi_{oo}^{-1}(z)$: IPSDM among $\mV_o$, $\varepsilon$,  $z=e^{j\omega},~\omega \in (-\pi,\pi]$\\
		\textbf{Output:} Matrices $\Im(\S(z))$ and $\Im(\L(z))$ 
		\label{alg:matrix_decomposition}
		\begin{algorithmic}[1]
		\State Set $\C=\Im\{\Phi_{oo}^{-1}(z)\}$ \State Initialize $(\widehat{\S}_{0},\widehat{\L}_{0})=(\C,\mathbf{0})$
			
			\ForAll{$t \in \{\eps,2\eps,\dots,1\}$}
			\State Solve the convex optimization \eqref{eq:convex_opti_t} and calculate ${diff}_t$ in \eqref{eq:diff_t}
			\EndFor
			\State Identify the three regions where $diff_t$ is zero and denote the middle region as $[t_1,t_2]$. 
			Pick a $t_0 \in [t_1,t_2]$ and the corresponding pair $(\hat{\S}_{t_0},\hat{\L}_{t_0})$.
\If{$deg_{max}(\hat{\S}_{t_0})inc(\hat{\L}_{t_0})<\frac{1}{12}$}

\State {$(\widehat{\S}(z),\widehat{\L}(z))=(\hS_{t_0},\hL_{t_0})$}		
\State {Return $(\widehat{\S}(z),\widehat{\L}(z))$ }
\EndIf 
\end{algorithmic}
\end{algorithm}

\section{Simulation results}
\label{sec:sim}
\begin{figure}[htbp]
\centering
\begin{subfigure}[t]{\linewidth}
\includegraphics[trim=150 250 80 170,clip, width=\textwidth]{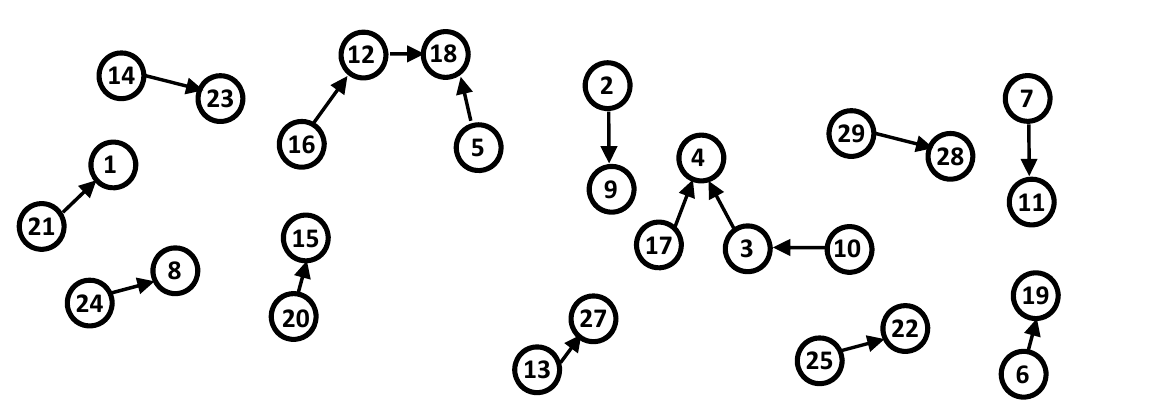}
\caption{The true LDIG, $\mG(\mV,\mE)$, of $(\H,\e)$.}
\label{fig:gen_graph}
\end{subfigure}
~
\begin{subfigure}[t]{\linewidth}
\includegraphics[trim=130 230 90 170,clip, width=\textwidth]{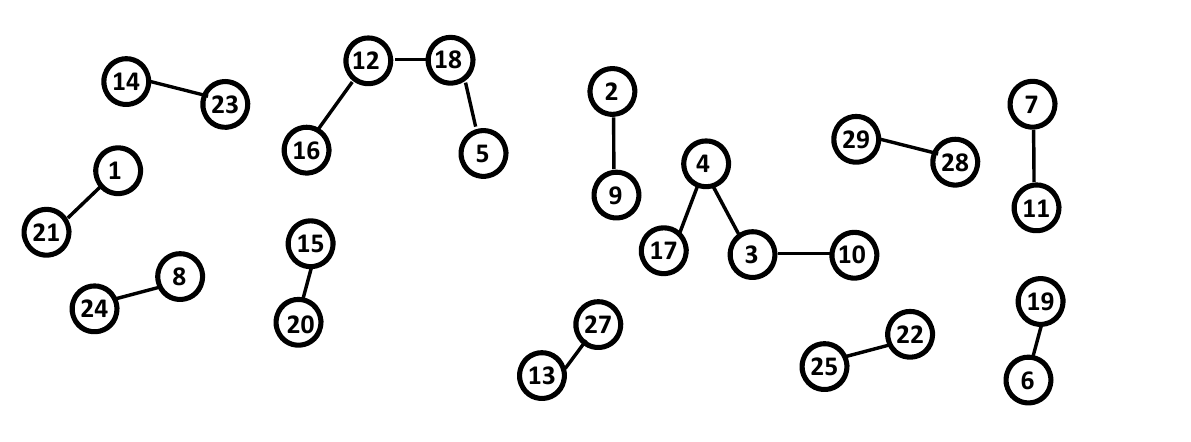}
\caption{The true topology, $\mT(\mV,\mE)$ of $(\H,\e)$.} 
\label{fig:top_gen_graph}
\end{subfigure}

\begin{subfigure}[t]{\linewidth}
\includegraphics[trim=130 250 90 140,clip, width=\textwidth]{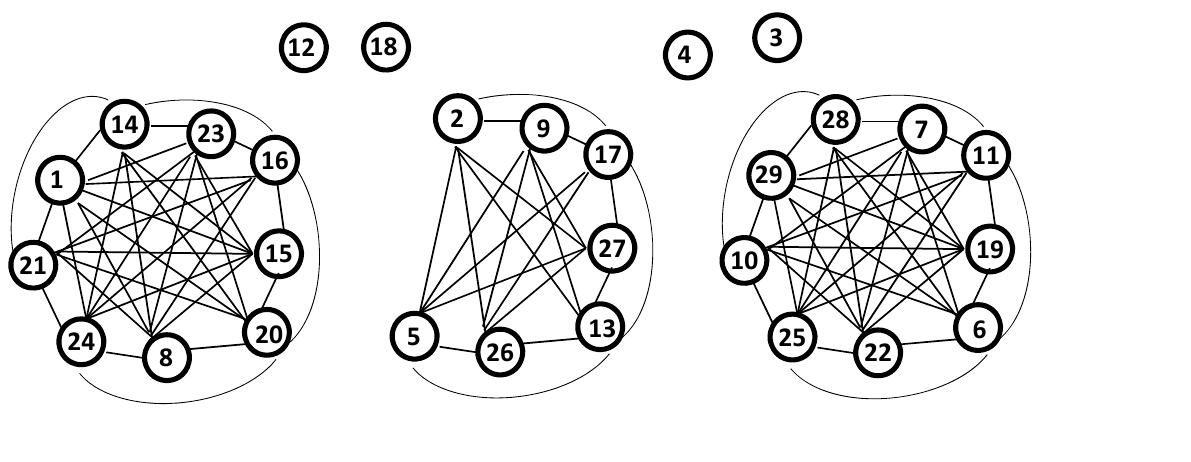}
\caption{Correlation graph, $\mG_c(\mV,\mE_c)$, of $(\H,\e)$;$M=3$} 
\label{fig:cliques}
\end{subfigure}

\begin{subfigure}[t]{\linewidth}
\includegraphics[trim=150 230 90 170,clip, width=\textwidth]{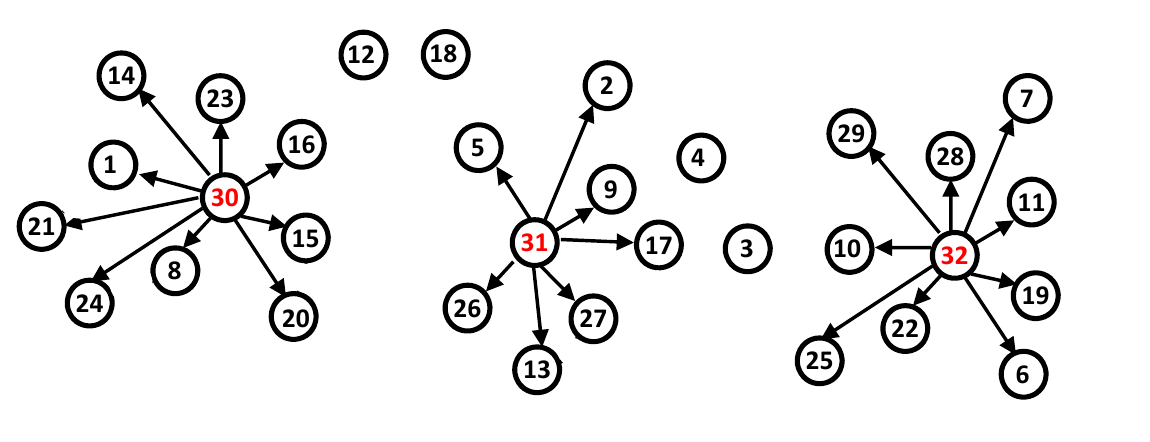}
\caption{Transformed correlation structure with latent nodes} 
\label{fig:gen_graph_latent_nodes}
\end{subfigure}
\begin{subfigure}[t]{\linewidth}
\includegraphics[trim=150 230 90 170,clip, width=\textwidth]{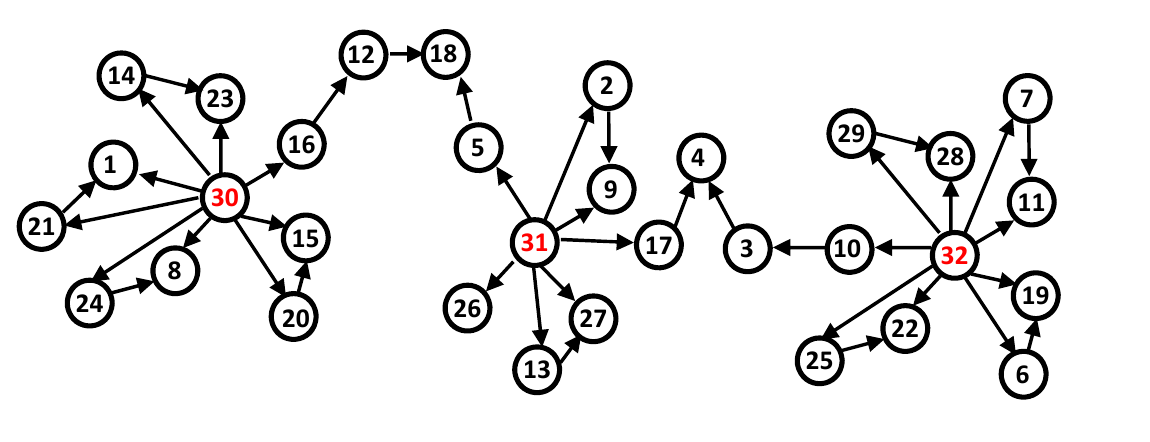}
\caption{Transformed graph, $\mG_t(\mV_t,\mE_t)$, of $(\bwH,\bwe)\in \mL_q(\H,\e)$.} 
\label{fig:complete_transformed_graph}
\end{subfigure}
\caption{Network structure} 
\label{fig:simulation_graphs}
\end{figure}

In this section, we demonstrate topology reconstruction of an LDIM $(\H,\e)$ with $\Phi_e$ non-diagonal, from $\Phi_\x^{-1}$, using the sparse$+$low-rank decomposition technique discussed in Section \ref{sec:topid_MD} for an affinely correlated network. Fig. \ref{fig:simulation_graphs}(a)-\ref{fig:simulation_graphs}(e) respectively depicts $\mG(\mV,\mE)$, $\mT(\mV,\mE)$, $\mG_c(\mV,\mE_c)$, $\mG(\mV_t,\mE_t\setminus \mE)$, and  $\mG_t(\mV_t,\mE_t)$ described in Section \ref{sec:G_ctoG_t-trans}. Simulations are done in Matlab. Yalmip \cite{yalmip} with SDP solver \cite{sdpt3} is used to solve the convex program \eqref{eq:convex_opti_t}.

For the simulation, we assume we have access to the true PSDM, $\Phi_\x$, of the LDIM of Fig. \ref{fig:simulation_graphs}(a). Here, $\Phi_\e$ is non-diagonal with the (unkown) correlation structure as shown in Fig. \ref{fig:simulation_graphs}(c).

\begin{figure}[htbp]
\centering
\includegraphics[trim=130 230 80 250,clip, width=.7\linewidth]{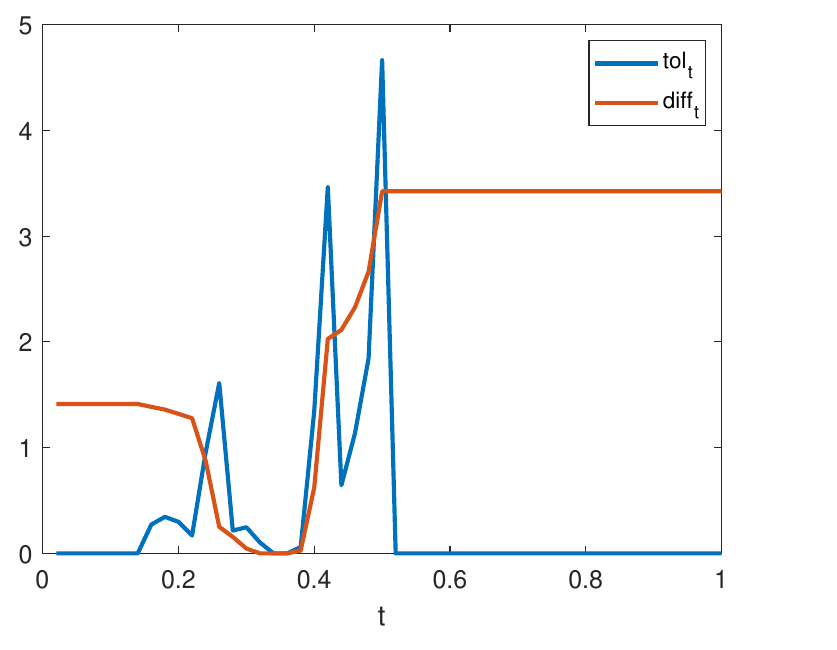}
\caption{tol$_t$ and diff$_t$ plots for the network of Fig. \ref{fig:simulation_graphs}(a) under affine correlation}
\label{fig:tol_diff}
\end{figure}

For the reconstruction, the imaginary part of the IPSDM, $\C=\Im\{\Phi^{-1}_\x(z)\}$ is employed in the convex optimization \eqref{eq:convex_opti_t} for $z=e^{2\pi/8}$.
Optimization \eqref{eq:convex_opti_t} is solved for all the values of $t \in [\epsilon,1]$, with the interval $\epsilon=0.01$. Notice that for $t=0$ $(\hS_t,\hL_t)=(\C,0)$. Fig. \ref{fig:tol_diff} shows the comparison of $tol_t$ and diff$_t$ versus $t$.  $(\hS_t,\hL_t)$ for $t=0.36$ is picked, which belongs to the middle zero region of diff$_t$ as described in \cite{doddi2019topology}.

${\mathbb I}_{\{\hS_t\neq \zero\}}$ returned the exact topology of Fig. \ref{fig:simulation_graphs}(b). From ${\mathbb I}_{\{\hL_t\neq \zero\}}$, by following Algorithms $2$ and $3$ in \cite{doddi2019topology},  $\mG_c(\mV,\mE_c)$ also is reconstructed, which matches Fig. \ref{fig:simulation_graphs}(c) exactly.

\begin{figure}[htbp]
\centering
\includegraphics[trim=70 200 40 200,clip, width=.7\linewidth]{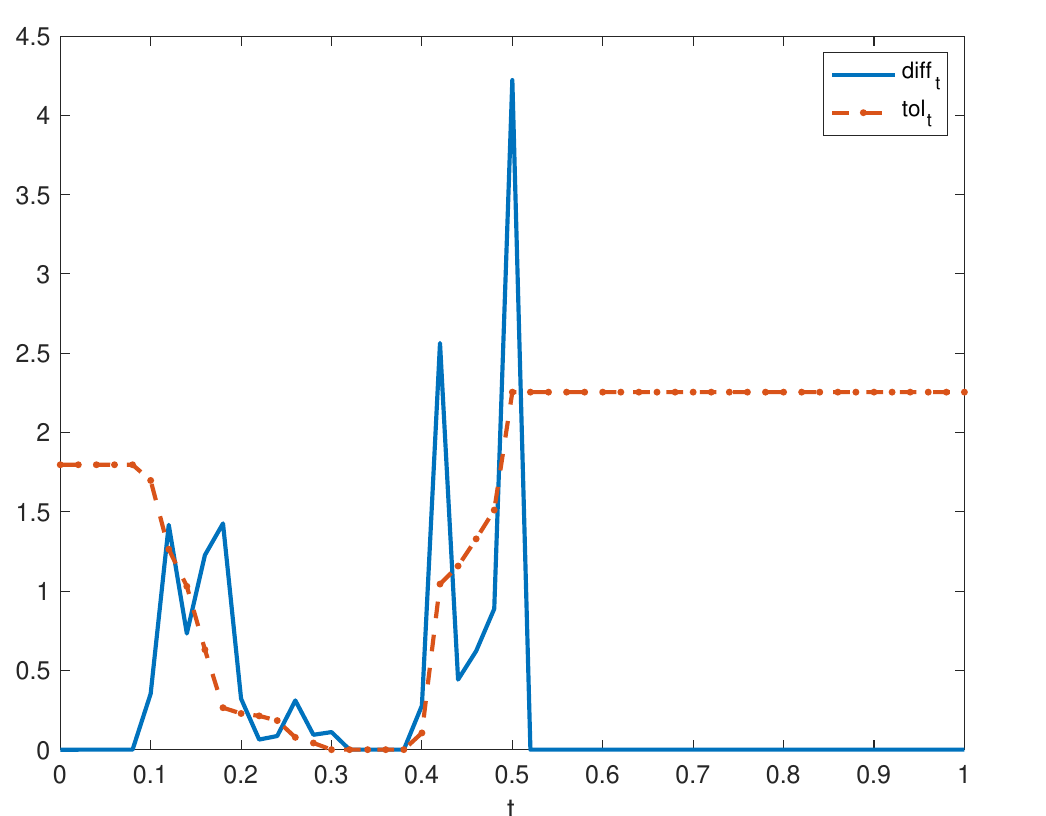}
\caption{tol$_t$ and diff$_t$ plots for the network of Fig. \ref{fig:simulation_graphs}(a) under polynomial correlation}
\label{fig:tol_diff_poly}
\end{figure}

\begin{figure}[htbp]
\centering

\begin{subfigure}[t]{\linewidth}
\includegraphics[trim=180 340 740 40,clip, width=\linewidth]{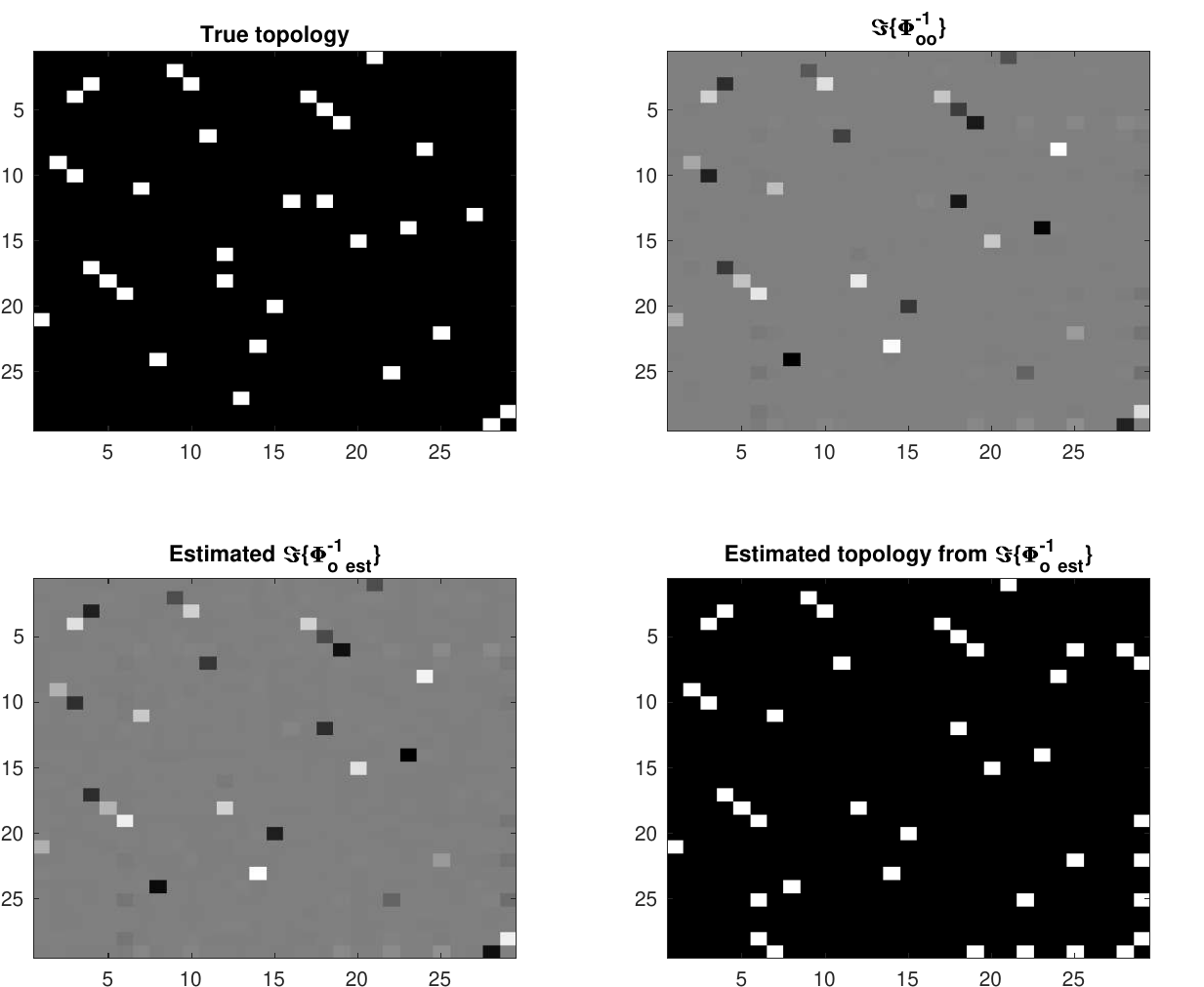}
\caption{The true topology and the topology estimated 
directly from $\widehat{\Phi}_o^{-1}(e^{j2\pi/5})$ for the network shown in Fig. \ref{fig:simulation_graphs}(a) under affine correlation with $N=10^6$ data samples. The figures show the entries of matrices of size $29 \times 29$. In the figures of topology, black denotes zero (edge absent) and white denotes 1 (edge present). In the gray scale images, gray denotes zero.}
\label{fig:matrix_IPSDM}
\end{subfigure}
~
\begin{subfigure}[t]{\linewidth}
\includegraphics[trim=800 40 120 40,clip, width=\linewidth]{matrices_full.pdf}
\caption{$\hS_t$ obtained by solving \eqref{eq:convex_opti_t} for  $\C=\Im\{\widehat{\Phi}_o^{-1}(e^{j2\pi/5})\}$. Topology is estimated from $\hS_t$ for $t=0.34$. In the figures of topology, black denotes zero (edge absent) and white denotes 1 (edge present). In the gray scale images, gray denotes zero.}
\label{fig:matrix_estimated_top}
\end{subfigure}
\caption{}
\label{fig:matrices}
\end{figure}


\subsection{Polynomial correlation}
\label{subse:simu_poly}
Here, the topology reconstruction of the network when the noise processes are polynomially correlated, is shown. For the simulation, consider the LDIG shown in Fig. \ref{fig:simulation_graphs}(a) and correlation graph of \ref{fig:simulation_graphs}(b). The noise processes are IID GP as shown in Section \ref{subsec:IID_Gauss} with $m=2$ and $p=3$. The entries of $\F$ are such that only the co-efficients corresponding to $y_2,y_5,$ and $y_9$ are non-zero, and the coefficients $F_{ij}=0$, for every $1\leq i\leq 29$, $1\leq j \leq 10,~ j\notin \{2,5,9\}$.

Figure \ref{fig:tol_diff_poly} shows the diff$_t$ and tol$_t$ plot of applying Algorithm \ref{alg:matrix_decomposition} with $\epsilon=0.01$. As shown in the plots, for $t\in[.28,.38]$, tol$_t$ is zero, which corresponds to the exact decomposition. Additionally, as mentioned in Lemma \ref{lem:diff_t}, diff$_t$ is zero in this interval. The support of $\hS_t$ for some $t\in[.28,.38]$ reconstructs the exact topology, which validates Theorem \ref{thm:s+L_top}. 

\subsection{Finite data simulation}
In this section, to evaluate the effect of finite data size, simulations are run on a synthetic data set based on the network shown in Fig. \ref{fig:simulation_graphs}(e). For the PSD estimation, Welch method \cite{stoica2005spectral} is used. Notice that the accuracy and the sample complexity of the estimation can be improved by employing advanced IPSDM estimation techniques from literature, for example see \cite{alpago2021scalable,zorzi_AR_hidden,ZORZI_Chiuso}. Fig. \ref{fig:matrices} shows the estimated results from a sample size of $10^6$; Fig. \ref{fig:matrices}(a) shows the true and the estimated IPSDM matrices. The fourth matrix of Fig. \ref{fig:matrices}(a) shows the topology estimated directly from $\Im\{\widehat{\Phi}_o^{-1}(e^{j2\pi/5})\}$, without decomposition. The estimation is done by hard thresholding. That is, the edge $(i,j)$ is detected if $[\Im\{\widehat{\Phi}_o^{-1}(e^{j2\pi/5})\}]_{i,j}$ is greater than a threshold, and not detected otherwise. The detection threshold is selected to obtain the minimum number of errors. 14 out of 16 edges are detected but 6 false positive edges are also detected, with the total error of 50\% (8 out of 16 edges). This shows that estimating the topology directly from $\Im\{\widehat{\Phi}_o^{-1}(e^{j2\pi/5})\}$ returns an undesirable number of errors.

 Towards the exact topology retrieval, the optimization \eqref{eq:convex_opti_t} is performed on  $\Im\{\widehat{\Phi}_o^{-1}(e^{j2\pi/5})\}$ to obtain sparse plus low-rank decomposition. Fig. \ref{fig:matrices}(b) shows the sparse part retrieved from the decomposition of $\Im\{\widehat{\Phi}_o^{-1}(e^{j2\pi/5})\}$ for various $t$ from $0$ to $0.5$. $\hS_t$ at $t=0.34$ is selected for estimating the topology. Thus, as illustrated by the above example,  with the approach proposed in the article it is possible to choose a threshold that yields 100\% detection without sacrificing the false alarm performance. 
 
 In order to demonstrate that the techniques proposed in this article do not degrade drastically with lesser number of samples, a simulation is run on 6000 samples. Employing the detection from $\Im\{\widehat{\Phi}_o^{-1}(e^{j2\pi/5})\}$ returned 48 false edges and missed one edge, thus a total of 49 errors. However, with the decomposition, detection from the support of $\hS_t$ at $t=.34$ detected 14 out of 16 edges and missed none, thus giving an error rate of 87.5\%. This confirms that the decomposition based reconstruction proposed by the article yield substantial advantages. The sample complexity analysis of the article's methods is open for future research.
 

\section{Conclusion}
\label{sec:conclusion}
In this article, the problem of reconstructing the topology of an LDIM with spatially correlated noise sources was studied. First, assuming affine correlation and the knowledge of correlation graph, the given LDIM was transformed into an LDIM with latent nodes, where the latent nodes were characterized using the correlation graph and all the nodes were excited by uncorrelated noises. For polynomial correlation, a generalization of the affine correlation, the latent nodes in the transformed LDIMs were excited using clusters of noise, where the noise clusters were uncorrelated with each other. Finally, using a sparse low-rank matrix decomposition technique, the exact topology of the LDIM was reconstructed, solely from the IPSDM of the true LDIM, when the network satisfied a sufficient condition required for the matrix decomposition. Simulation results are provided that verify the theoretical results.

\appendix
\section{Proof of Lemma \ref{lem:clique_G_c-vs-latent_node}}
\label{app:clique_G_c-vs-latent_node}

Let $([\H,\F],\bwe) \in \mL(\H,\e)$ and let $i,j\in[n]$. By definition of $\mL(\H,\e)$, $e_i=\we_i+\F_i \bwe_h$, where $\F_i$ denotes the $i^{\text{th}}$ row of $\F$, for any $i \in [n]$. 
To prove the only if part, suppose that 
$\Phi_{e_ie_j} 
\neq 0$. By definition, $e_ie_j =  (\we_i+\F_i \bwe_h)(\we_j+\F_j \bwe_h)$.
Then, $ \Phi_{e_ie_j}=\F_i \Phi_{\bwe_h}\F_j^*$
$=\sum_{k}F_{ik}F_{jk}^* \Phi_{\we_{h_{k}}}$, since $\we_i$ and $\we_j$ are uncorrelated for $i,j \in [n+L]$.
 Thus, $\sum_{k}F_{ik}F_{jk}^*\Phi_{\we_{h_{k}}} \neq 0$,  which implies that there exists a $k$ such that $F_{ik}\neq 0$ and $F_{jk} \neq 0$. In other words, there exists a latent node $h_k$ in the corresponding LDIG of $([\H,\F],\bwe)$ such that $i,j \in Ch(h_k)$. 

$\Leftarrow$ 
Let $i,j \in [n]$ such that $\Phi_{e_ie_j}=0$. Then, from the proof of only if part, $0=\F_i \Phi_{\bwe_h}\F_j^* =\sum_{k}F_{ik}F_{jk}^* \Phi_{\we_{h_{k}}}, $
which implies that $F_{ik}=0$ or $F_{jk}=0, \ \forall k$, except for a few pathological cases that occur with Lebesgue measure zero. We ignore the pathological cases here. Hence, there does not exist any latent node $h$ such that both $i,j \in Ch(h)$.\hfill \hfill \qed

The following result shows that if a subgraph, $\mG^\ell(\mV^\ell,\mE^\ell)$, of the correlation graph, $\mG_c(\mV,\mE_c)$, forms a maximal clique in $\mG_c$, then for any transformed LDIG in $\mL(\H,\e)$, the set of nodes in $\mV^l$ is equal to the set of children of some latent nodes in the LDIG.

\section{Proof of Lemma \ref{lem:Q_leq_L}}
\label{app:lem_Q_leq_L}
The following lemma is useful in proving Lemma \ref{lem:Q_leq_L}.
\begin{lem}
\label{lem:clique-vs_lat-nodes}
Let $(\H,\e)$ be an LDIM defined by \eqref{eq:Corr_TF_model} which satisfies Assumption \ref{ass:affine_corre}, and let $\mG_c(\mV,\mE_c)$ be the correlation graph of the exogenous noise sources, $\e$. Suppose that $\mG^\ell(\mV^\ell,\mE^\ell)\subseteq \mG_c(\mV,\mE_c)$ is a maximal clique with $|\mV^\ell|>1$. Then, for every LDIG $([\H,\F],\bwe) \in  \mL(\H,\e)$, there exist latent nodes $h_1^\ell,\dots,h^\ell_{k_\ell}$ such that	
	\begin{align}
	\label{eq:V^l_condition}
	\mV^\ell&=\bigcup_{i=1}^{k_\ell}Ch(h^\ell_i) \text{ and }
	\mE^\ell= \bigcup_{i=1}^{k_\ell}\mE_{\ell,i},
	\end{align} 
	where  $\mE_{\ell,i}:=\{(k,j):k,j \in Ch(h_i^\ell)\}$.

	In particular, for any latent node $h$ in the LDIG $([\H,\F],\bwe)$, $Ch(h) $ forms a clique (not necessarily maximal) that is restricted to $\mG^\ell(\mV^\ell,\mE^\ell)$ in $\mG_c$.
\end{lem}
\textbf{Proof:} 
Let $\mV^\ell \subseteq [n]$, $ |\mV^l|>1$, such that $\mV^l$ forms a clique in $\mG_c$. Lemma \ref{lem:clique_G_c-vs-latent_node} showed that, for any $i,j \in \mV^l$, there exists a latent node $h$ such that $i,j \in Ch(h)$ in the LDIG, $\mG$, of $(\bwH,\bwe)$. Since this is true for any pair $i,j \in \mV^l$,
there exists a minimal set of latent nodes $\mathcal H^l:=\{h^l_i\}_{i=1}^{k_l}$, $k_l \in \mathbb{N}\setminus \{0\}$, s. t. for any $i,j \in \mV^l$, we have $i,j \in Ch(h_p^l)$ for some $h_p^l \in \mathcal H$, in $\mG$. Hence, $\mV^l\subseteq \bigcup_{i=1}^{k_l}Ch(h^l_i)$. Similarly, $i,j \in Ch(h_p^l)$ implies $(i,j) \in \mE_{l,p}$. Therefore, $\mE^l\subseteq \bigcup_{i=1}^{k_l}\mE_{l,i}$.

Next, we prove that all the children of a given latent node belong to a single (maximal) clique, which proves $\mV^l\supseteq \bigcup_{i=1}^{k_l}Ch(h^l_i)$ and $\mE^l\supseteq \bigcup_{i=1}^{k_l}\mE_{l,i}$.
Let $h \in \mathcal H^l$ be a latent node in the LDIG of $(\bwH,\bwe)$ and suppose $i,j \in Ch(h)$. Then, from the definition of $(\bwH,\bwe)$, there exists $\ell$ such that $F_{i\ell} \neq 0$ and $F_{j\ell} \neq 0$, and hence $\Phi_{e_ie_j} = \F_i \Phi_{\bwe_h}\F_j^* \neq 0$ a.e. Thus, $(i,j) \in \mE_c$, excluding the pathological cases. Notice that this is true for any $i,j \in Ch(h)$. Therefore, $Ch(h) \subseteq \mV^{\ell}$ forms a clique (not necessarily maximal) in $\mG_c(\mV,\mE_c)$. Since this is true for every $h\in \mathcal{H}^l$, $\mV^l\supseteq \bigcup_{i=1}^{k_l}Ch(h^l_i)$. The similar proof shows that $\mE^l\supseteq \bigcup_{i=1}^{k_l}\mE_{l,i}$.
\hfill \hfill \qed

The proof of Lemma \ref{lem:Q_leq_L} follows directly from \eqref{eq:V^l_condition} and by noting that $k_\ell \geq 1$\hfill \hfill \qed

\section{Proof of Theorem \ref{thm:L=Q}}
\label{app:lem_L=Q}
\begin{lem}(Pigeonhole principle): The pigeonhole principle states that if $n$ pigeons are put into $m$ pigeonholes, with $n>m$, then at least one hole must contain more than one pigeon.
\end{lem}
 We use pigeonhole principle and Lemma \ref{lem:clique-vs_lat-nodes} to prove this via contrapositive argument. Recall that the number of latent nodes $L$ is equal to the number of cliques $M$. Suppose there exists a clique $\mG^l(\mV^l,\mE^l)$ such that, for some $i,j \in \mV^l$, there does not exist a latent node $h \in (\bwH,\bwe)$ with $i,j \in Ch(h)$. By Lemma \ref{lem:clique-vs_lat-nodes}, there exist latent nodes $h,h'$ such that $i \in Ch(h)$ and $j \in Ch(h')$. By Lemma \ref{lem:clique-vs_lat-nodes} again, all the children of $h$ are included in a single clique. That is, $Ch(h) \subseteq \mV^l$ and $Ch(h') \subseteq \mV^l$,  since $i,j \in \mV^l$. Then, excluding $\mV^l$, $h$, and $h'$, we are left with $M-1$ cliques and $L-2=M-2$ latent nodes. Applying pigeonhole principle, with $M-1$ pigeons (cliques) and $L-2$ holes (latent nodes), there would exist at least one latent node $k$ with $Ch(k)$ belonging to two different maximal cliques, which is a contradiction of Lemma \ref{lem:clique-vs_lat-nodes}.
\hfill \hfill \qed

\section{Proof of Proposition \ref{prop:topology_uniqueness}}
\label{app:proof_prop_top_uniqueness}
Let $T_1, \ T_2 \in \{0,1\}^{(n+L) \times (n+L)}$, $T_1 \neq T_2$, be the topologies of two distinct transformations $(\bwH^1,\bwe^1)$ and $(\bwH^2,\bwe^2)$ respectively. Without loss of generality, let $(i,j)$ be such that $[T_1]_{ij} \neq 0 $ and $[T_2]_{ij} = 0$. By the definition of $([\H,\F],\bwe)$, if $i \leq n$ and $j \leq n$, then $[T_1]_{ij}=[T_2]_{ij}=\1_{\{\{H_{ij}\neq0\} \cup \{H_{ji}\neq0\} \}}$. If $i \leq n$ and $j \geq n$, then $i$ is an observed node and $j$ is a latent node. From Lemma \ref{lem:clique_G_c-vs-latent_node}, $F^1_{ij}=F^2_{ij}=0$ if and only if $j \notin Pa(i)$. Thus, $[T_1]_{ij}=[T_2]_{ij}$, which is a contradiction, since both cannot be true. Similar contradiction holds if $i \geq n$ and $j \leq n$. If $i,j>n$, then $[T_1]_{ij}=[T_2]_{ij}=0$ since $Pa(i)=Pa(j)=\emptyset$.  Thus, the assumption leads to a contradiction, which implies that $T_1=T_2$. \hfill \hfill \qed

\section{Proof of Proposition \ref{prop:Gaussian_block_diagonal}}
\label{app:prop_Gaussian_block_diagonal}
 Consider a pair of monomials $y_k, y_l$ with $y_k= \v(0)^\alpha$ and $y_l= \v(0)^\beta$. For notational convenience, the index $0$ is omitted. Then, $\expectation{y_ky_l}=\expectation{\v^{\alpha+\beta}}$ $\displaystyle=\prod_{i=1}^m\expectation{\v_i^{\alpha_i+\beta_i}}$. By \eqref{eq:moments_Gaussian}, $\expectation{y_ky_l}\neq 0$ if and only if  $\alpha_i+\beta_i$ is even, $\forall i\in[m]$. Suppose $\alpha_i$ is odd. Then, $\beta_i$ must be odd. Similarly, $\beta_i$ must be even if $\alpha_i$ is even, $\forall i \in [m]$. 

Define an element-wise boolean operator, $\mB: \mathbb N^m \mapsto \{0,1\}^m$ such that for $\u=\mB(\alpha)$, $u_i=0$ if $\alpha_i$ is odd and $u_i=1$ if $\alpha_i$ is even. Then, for $y_k= \v^\alpha$ and $y_l= \v^\beta$, $\expectation{y_ky_l}\neq 0$ if and only if $\mB(\alpha)=\mB(\beta)$. Group the monomials with the the same odd-even pattern into one cluster. Since the total different values that $\mB(\cdot)$ can take is $2^m$, there are $2^m$ distinct clusters that are uncorrelated with each other.



Reorder $\y$ by grouping the monomials belonging to the same cluster together to obtain $\ty$, similar to the $(n,p)=(2,3)$ example in Section \ref{subsec:IID_Gauss}. Then, $\Phi_{\ty}$ is a block diagonal matrix with $2^m$ blocks for $m$ variable polynomials, where each diagonal block corresponds to one particular pattern of $\{Odd,Even\}^m$. \hfill \hfill \qed

\section{Proof of Lemma \ref{lem:poly_clique_G_c-vs-latent_node}}
\label{app:poly_clique_G_c-vs-latent_node}
Let $(\H,\F,\bwe) \in \mL(\H,\e)$ and let $i,j\in[n]$. By definition of $\mL(\H,\e)$, $e_i=\we_i+\F_i \bwe_h$, where $\F_i$ denotes the $i^{\text{th}}$ row of $\F$, for any $i \in [n]$. 
To prove the only if part, suppose that $\Phi_{e_ie_j} 
\neq 0$. By definition, $e_ie_j =  (\we_i+\F_i \y)(\we_j+\F_j \y)$.
Since $\bwe_i,\bwe_j$, and $\bwe_h$ are uncorrelated, and $\Phi_{\y}$ is block diagonal (Proposition \ref{prop:WSS_blockdiag}),$$ \Phi_{e_ie_j}=\F_i \Phi_{\y}\F_j^*=\sum_{c=1}^{2^m}\sum_{k_1,k_2\in \mC_c}F_{ik_1}F_{jk_2}^* \Phi_{\y_{k_1}\y_{k_2}}\neq0.$$ 
Thus, there exists a $c$ such that $F_{ik_1}\neq 0$ and $F_{jk_2} \neq 0$ for some $k_1,k_2\in \mC_c$. That is, there exists a cluster $c$ such that $i,j \in Ch(c)$.

$\Leftarrow$ Let $i,j \in [n]$ such that $\Phi_{e_ie_j}=0$. Then, from the proof of only if part, {$\sum_{c=1}^{2^m}\sum_{k_1,k_2\in \mC_c}F_{ik_1}F_{jk_2}^* \Phi_{\y_{k_1}\y_{k_2}}=0$}. That is, $F_{ik}=F_{jk}=0, \ \forall k \in \mC_c, ~\forall c\in [2^m]$, except for a few pathological cases that occur with Lebesgue measure zero. We ignore the pathological cases here. Hence, there does not exist any cluster $c$ such that both $i,j \in Ch(c)$.\hfill \hfill \qed

\section{Proof of Theorem \ref{thm:prop_poly_clique-vs_lat-nodes}}
\label{app:thm_poly_clique-vs_lat-nodes}
The proof follows similar to the proof of Lemma \ref{lem:clique-vs_lat-nodes}, by replacing latent nodes with latent cluster, as in the proof of Lemma \ref{lem:poly_clique_G_c-vs-latent_node}.\hfill \hfill \qed
\section{Proof of Theorem \ref{thm:s+L_top}}
\label{app:proof_thm_S+L_top}
\vspace{-10pt} As shown in Lemma \ref{lem:suff_cond}, if $deg_{max}(\Im\{\S\})inc(\Im\{\L\})<1/12$, then for the appropriately selected $t$, the convex program \eqref{eq:convex_opti_t} retrieves $\Im\{\S\}$ and $\Im\{L\}$ when $\C=\Im\{\S\}+\Im\{\L\}$. If any one of the LDIMs  $(\bwH,\e) \in \mL_q(\H,\e)$ satisfies this condition, then the imaginary part of $\hS_t=(\I_n-\H^{*})\Phi_{\we_o}^{-1}(\I_n-\H)$ returns the topology among the observed node of $(\bwH,\e)$, by Lemma \ref{lem:S_top_obs_rec}. The theorem follows from Remark \ref{remark:P1=P2} and Lemma \ref{lem:clique-vs_lat-nodes}. \hfill \hfill \qed

\begin{ack}                               
The authors acknowledge the support of NSF for supporting this research through the project titled "RAPID: COVID-19 Transmission Network Reconstruction from Time-Series Data" 
under Award Number 2030096.  
\end{ack}

\bibliographystyle{plain}        

\end{document}